\PassOptionsToPackage{square,numbers,compress}{natbib}
\documentclass[subscriptcorrection,upint,varvw,barcolor=black,
mathalfa=cal=euler,balance,hyphenate,french]{asmejour}

\PreprintString{ACCEPTED MANUSCRIPT} %

\usepackage{amsmath} %
\usepackage{amsfonts} %
\usepackage{amsopn} %

\DeclareMathOperator*{\argmin}{arg\,min}
\usepackage{mathrsfs} %
\usepackage{euscript} %
\usepackage{mathtools}

\usepackage{graphicx} %

\usepackage[strings]{underscore} %
\usepackage[strict=true]{csquotes} %
\usepackage{cleveref}           %
\crefname{figure}{Fig.}{Figs.} %

\usepackage[normalem]{ulem}

\usepackage[switch]{lineno}

\newcommand{\tmp}[1]{\textcolor{white!40!black}{#1}} %

\hypersetup{%
	pdfauthor={Taiwo A. Adebiyi}, %
	pdftitle={DT Civil Review},   %
}

\JourName{Computing and\\ Information Science in Engineering}
\PaperYear{}

\begin{document}

\SetAuthorBlock{Taiwo A. Adebiyi}{Department of Civil and Environmental Engineering,\\
   University of Houston,\\
   4226 Martin Luther King Boulevard,\\
   Houston, TX, 77204-4003  \\
   email: taadebi2@cougarnet.uh.edu} 

\SetAuthorBlock{Nafeezat A. Ajenifuja}{Department of Civil and Environmental Engineering,\\
   University of Houston,\\
   4226 Martin Luther King Boulevard,\\
   Houston, TX, 77204-4003  \\
   email: naajenif@cougarnet.uh.edu}

 \SetAuthorBlock{Ruda Zhang\CorrespondingAuthor}{%
Assistant Professor\\
Department of Civil and Environmental Engineering, \\
University of Houston, \\
4226 Martin Luther King Boulevard, \\
Houston, TX, 77204-4003  \\
email: rudaz@central.uh.edu
}

\title{Digital Twins and Civil Engineering Phases:\\ Reorienting Adoption Strategies}


\begin{abstract}
Digital twin (DT) technology has received immense attention over the years due to the promises it presents to various stakeholders in science and engineering. As a result, different thematic areas of DT have been explored. This is no different in specific fields such as manufacturing, automation, oil and gas, and civil engineering, leading to fragmented approaches for field-specific applications. The civil engineering industry is further disadvantaged in this regard as it relies on external techniques by other engineering fields for its DT adoption. A rising consequence of these extensions is a concentrated application of DT to the operations and maintenance phase. On another spectrum, Building Information Modeling (BIM) is pervasively utilized in the planning/design phase, and the transient nature of the construction phase remains a challenge for its DT adoption. In this paper, we present a phase-based development of DT in the Architecture, Engineering, and Construction industry. We commence by presenting succinct expositions on DT as a concept and as a service, and establish a five-level scale system. Furthermore, we present separately a systematic literature review of the conventional techniques employed at each civil engineering phase. In this regard, we identified enabling technologies such as computer vision for extended sensing and the Internet of Things for reliable integration. Ultimately, we attempt to reveal DT as an important tool across the entire life cycle of civil engineering projects, and nudge researchers to think more holistically in their quest for the integration of DT for civil engineering applications.
\end{abstract}

\date{}

\maketitle %

\section{Introduction}

The idea of Digital Twin (DT) has evolved as a leading technological idea of the fourth industry revolution, with Gartner research firm predicting the idea as one of the top ten most promising technology trends over the next ten years \cite{Tao2017}. We do not find such estimation surprising as DT has successfully evolved across several industries, especially in manufacturing \cite{Liu2019,Qi2018}, aviation \cite{Guo2018,Tuegel2011}, and oil and gas sectors \cite{Cameron2018,Zhang2020}. For example, leading technological firms such as General Electric (GE), Siemens, British Petroleum (BP), and Airbus, have implemented DTs for optimized production \cite{Pan2021,Tao2019}. As already stated by several researchers \cite{Enzer2019,Lamb2019,Angjeliu2020}, such a level of DT adoption has not been witnessed in the civil engineering industry. Different reasons have been highlighted for this slow pace which include the perceived rigidity of the civil engineering profession and the complexity of its projects. Recalling that every civil engineering project must evolve across three distinct but sequential phases including planning/design, construction, and operation and maintenance, the direct adoption of DT from other engineering disciplines will be limiting \cite{Angjeliu2020}. Thus, the transient nature of civil engineering projects and the longevity of their built structures compels a simplified representation of DT with increasing complexities at consecutive phases. Pregnolato et al.~\cite{Pregnolato2022} also highlighted that the slow rate of change of civil engineering assets makes real-time twinning---a critical part of DT---challenging.

Currently, most civil-engineering-based attempts for DT creation rely on external techniques by other engineering fields such as manufacturing and aviation industries, birthing an indirect consequence of a segmented application of DT to the operation and maintenance phase. Torzoni et al.~\cite{Torzoni2024} extended a probabilistic graphic model already applied to an unmanned aerial vehicle originally formulated by Kapteyn et al.~\cite{Kapteyn2021} to develop a predictive DT for the health monitoring and management planning of existing structures. While several works of literature \cite{Jones2020,Rasheed2020, Wagg2020,Kritzinger2018} exist on the review of DT as a concept and its values to engineering systems and operations, such review-based works are practically non-existent in the civil engineering sector when compared to parallel industries. In a trend analysis study by Lamb \cite{Lamb2019}, just an estimate of six percent of published DT-related literature since 2016 is directly linked to the built environment, with about fifty percent of the literature connected to the manufacturing sector. Furthermore, only an extremely few number of publications focused on the general overview of DT for holistic civil engineering applications, as most published works are inclined towards DT as an optimized system for structural health monitoring. 

Arup \cite{Arup2019} in their review-based report on DT and its significance for the civil engineering profession established a metric system for scaling and classification, and also its values through case studies analysis. Pregnolato et al.~\cite{Pregnolato2022} reviewed the state-of-the-art DTs in civil engineering which mostly highlights their use on already constructed structures, a brief review on BIM and DT, and a proposed DT workflow process for bridge management. Despite the title: ``DT framework in civil engineering'', Torzoni et al.~\cite{Torzoni2024} in their recent publication merely extended existing mathematical models for SHM. Although Jiang et al. \cite{Jiang2021} discussed DT and civil engineering phases, the authors solely reported possible DT applications across these phases, focusing less on combining conventional civil engineering techniques and technologies with recent computational advances for DT creation. It is thus crucial to consider what DT means to civil engineering at every phase of its project life cycle, clarifying their distinctiveness and connectedness which is the crux of this paper. Importantly, we attempt to reveal DT as an important system across every civil engineering phase and nudge researchers to think more holistically in their quest for the integration of DT for civil engineering applications.

We commence this paper with succinct clarifications on DT definitions and phase-based conceptual overviews, classification, and scales in \Cref{sec: 2}. We then present a general review of DT and civil engineering phases in \Cref{sec: 3}, setting the tone for the next three sections. Sections \ref{sec: 4}, \ref{sec: 5}, and \ref{sec: 6} thoroughly explore each phase by introducing state-of-the-art tools and techniques that are conventionally applied, reporting demonstrated works, and developing concepts/frameworks for phase-based DT applications. Through our review-based methodologies, we generate insights and future research directions for holistic civil engineering applications, as we clearly show that all the tools needed for DT realizations are already available. We then conclude by summarizing the aforementioned expositions in \Cref{sec: 7}.  

\section{DT Conceptual Model Developments, Classification, and Scale}
\label{sec: 2}
We build on already existing views to reveal the central concepts of DT. Through our extensive reviews, we notice the definition of DT across two spectrums, i.e., in terms of development and in terms of service or an integration of both. As presented in Table~\ref{tab:1}, we notice that academic-based definitions are more inclined toward DT developments while industry-based definitions are inclined toward DT services (i.e., values created). Several works \cite{Jones2020,Arup2019,Rasheed2020,Kritzinger2018,Committee2020} have been done to resolve the varying definitions of DT, and as such, we will not follow suits in this regard but rather present the mandatory components of DT with emphasis on their distinguishing features and minimal services, leading to some classifications.

In the simplest possible form, we view DT as an advancement to existing technologies for the virtual representation of a physical system, with the added possibilities of real-time bi-directional data exchange, feedback, and control, giving way to several applications. We note that this view is sufficient for our expositions on DT development for the design and construction phases but lacks technical adequacy for the operational and maintenance phases of civil engineering projects. For this, we adopt Kapteyn et. al's \cite{Kapteyn2021} conceptual view of DT and its corresponding physical twin as two-way tightly coupled dynamical systems evolving through their respective state spaces. Through this view, probabilistic formulations are enabled: a fundamental part of uncertainty quantification (UQ) which we strongly note to be a key aspect of DT development and service.

\begin{figure}[h]
    \centering
    \includegraphics[width=0.5\textwidth]{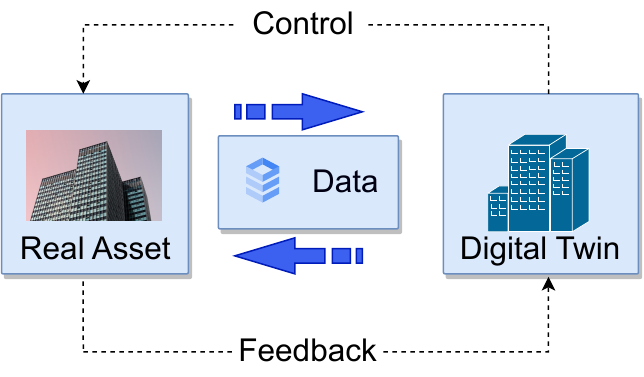}
    \caption{DT conceptual view. \textnormal{\textit{Requisite components include a virtual representation of a physical asset, real-time bi-directional data exchange, feedback, and control.}}}
    \label{fig:DT_Concept}
\end{figure}

At the bare minimum, a DT system should have a corresponding physical twin, bi-directional real-to-virtual connection, real-time data acquisition and update (feedback), and control i.e., being able to influence the actions of the physical twin (Figure \ref{fig:DT_Concept}).
To quantify the performance of DT, a reward metric is usually established \cite{Kapteyn2021}, which can also be termed service. To re-emphasize the requisite need for UQ, we need to debunk the perceptions of DT as being indistinguishable from its physical twin or as an exact mirror/replica of the same. As rightly pointed out by Arup \cite{Arup2019}, DT should be relevant abstractions of the physical twin. As opposed to the literal definition of a twin, modeling the entire physical state (if possible) of an asset will be computationally intractable, and as such DT will attempt to represent the corresponding digital state with a high level of complexity but simple enough for computational efficiencies \cite{Kapteyn2021}. Based on this, representative errors that occur across every component of DT must be accounted for to ensure reliability, wherein UQ is a critical recipe.

The varying conceptual views of DT have led to different types of classifications in the body of literature, two of which will be of importance for this work. First is a hierarchical classification based on the level of integration given by Kritzinger et al.~\cite{Kritzinger2018}. It starts with the \textit{Digital Model} which is simply a digital representation of an asset without automated data exchange between the physical asset and digital object. A further improvement on the digital model to include an automated one-way data flow is termed \textit{Digital Shadow}. Ultimately, a DT ensues when there is an automated bi-directional data exchange with its corresponding physical asset. This classification will particularly become important in DT's relation to BIM for civil engineering projects. Another is the service-based classification reported by Elfarri et al.~\cite{Elfarri2023} comprising DT, Digital sibling, and Digital threads. \textit{Digital Sibling}, also reported by \cite{Rasheed2020} does not necessarily run in real-time but is used for scenario-based (what-if) analysis for risk assessments, and \textit{Digital Threads} utilize overtime DT predictions for designing the next generation of assets. Having established the fundamental concepts of DT and classifications, we now present a scaling system to categorize the realizations of DT.

\begin{table*}[ht]
\caption{Academic and industry definitions of DT (Inspired by \cite{Arup2019}).\label{tab:1}}
\centering{%
\begin{tabular}{  p{3.7cm}   p{9.3cm}  p{3cm}}
\toprule
References & Definition & Spectrum \\
\midrule
Grieves (Academia) &  The digital twin is a set of virtual information constructs that fully describes a potential or actual physical 
manufactured product from the micro (atomic level) to the macro (geometrical level) \cite{Grieves2017digital}. In a further exposition, Grieves highlighted DT to consist of a physical product, a virtual representation, and bi-directional data connections for information exchange \cite{Grieves2014digital}. & Service + Development \\ [1.7cm]
IBM (Industry) & A digital twin is a virtual representation of an object or system that spans its lifecycle, is updated from real-time data, and uses simulation, machine learning, and reasoning to help decision-making \cite{IBM2024}. & Service + Development  \\ [1cm]
Rasheed et al. (Academia) & A digital twin is defined as a virtual representation of a
physical asset enabled through data and simulators for
real-time prediction, optimization, monitoring, controlling, and improved decision-making \cite{Rasheed2020}. & Service + Development  \\ [1cm]
AIAA (Academia + Industry) & A DT is a set of virtual information constructs that mimic the structure, context, and behavior of an individual/unique physical asset, or a group of physical assets, is dynamically updated with data from its physical twin throughout its life cycle and informs decisions that realize value. They advanced the essential elements of DT to comprise a model, asset, and data/information connectedness between the two \cite{Committee2020}. & Service + Development \\ [1.7cm]

General Electric (Industry) & A digital twin is a living model that drives a business outcome \cite{GE2019} & Service \\ [0.5cm]

Siemens (Industry) & A digital twin is a virtual representation of a physical product or process, used to understand and predict the physical counterpart’s performance characteristics.  \cite{Siemens2024} & Service \\ [0.9cm]

Glaessgen \& Stargel (Academia) & DT is an integrated multiphysics, multiscale, probabilistic simulation of an as-built system that uses the best available physical models, sensor updates, history, etc., to mirror the life of its corresponding twin \cite{Glaessgen2012}. & Service + Development \\
\bottomrule
\end{tabular}
}%
\end{table*}

\subsection{Scaling Metric for DT}

Arup \cite{Arup2019} developed a five-level scale quantifying the present and probable future capabilities of DT based on four metrics which include \textit{Autonomy, Intelligence, Learning, and Fidelity}. These four metrics must be individually evaluated from level one (1) to level five (5), and their cumulation results in a 5-level scale for DT as presented in Table \ref{tab: 2}.  Elfarri et al. \cite{Elfarri2023} reported a DT capability level scale in six (6) tiers---ranging from 0 to 5, with each tier characterized by their increasing level of complexities and functionalities, also presented in Table \ref{tab: 2}. We combined these scaling systems with our conceptual views of DT to introduce a corresponding five-level scale system. Precisely, a level-1 DT only involves digital representations of its respective asset system; an improvement to level-2 includes real-time monitoring, level-3 further includes predictive capability, level-4 permits interaction or learning from the twin data and environment, and level-5 comprises all of the aforementioned with a distinctive capacity for autonomous control. It should be noted that we consider the zeroth-level of \cite{Elfarri2023}'s scale to be negligible without loss of generality.

\begin{table*}[t]
\caption{Scaling of DT. \label{tab: 2}}
\centering{%
\begin{tabular}{  p{1cm}   p{5cm}  p{4.5cm} p{5cm}}
\toprule
Levels & Arup \cite{Arup2019} & Elfarri \cite{Elfarri2023} &  Current Work  \\
\midrule
1 & A digital model of an asset, which lacks intelligence, learning, and autonomy & Descriptive CAD models with real-time data acquisition & Digital representation of an asset \\ [0.5cm]
2 & Digital model with some capacity for feedback and control & Models with diagnostic possibilities &  Digital representation + real-time monitoring \\ [0.5cm]
3 & Digital model with capacity for predictive maintenance, analysis, and insights & 
Models with predictive capabilities & Digital representation + real-time monitoring + prediction \\ [0.5cm]
4 & A digital model with the capacity to learn from its asset data and surrounding environments & Models with prescriptive (i.e., providing recommendations) capabilities & Digital representation + real-time monitoring + prediction + interaction \\ [0.5cm]

5 & Digital models with several capabilities, and fully autonomous. & Models with autonomous capabilities  & Digital representation + real-time monitoring + prediction + interaction + autonomous control \\ [0.5cm]

\bottomrule
\end{tabular}
}%
\end{table*}

\section{Digital Twin and Civil Engineering Phases}
\label{sec: 3}
\begin{figure}[h]
    \centering
    \includegraphics[width=0.5\textwidth]{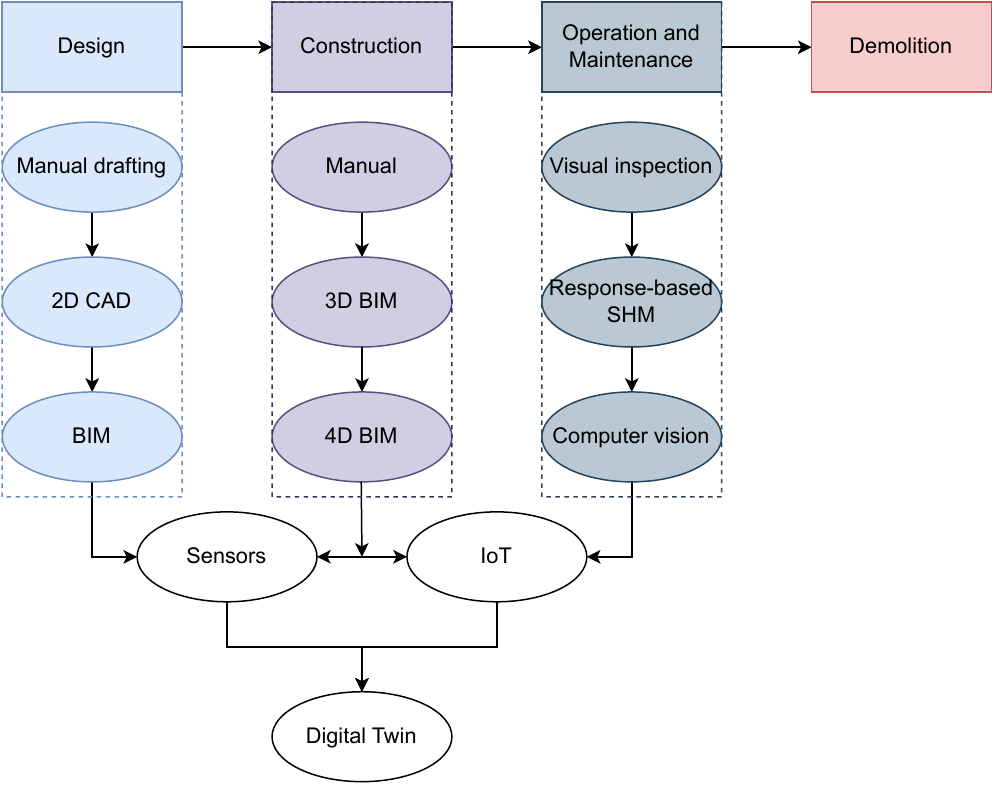}
    \caption{Civil engineering phases and evolution of enabling technologies}
    \label{fig: civil engineering phases}
\end{figure}

``Construction 4.0'' is the corresponding terminology of the fourth industrial revolution (Industry 4.0) in the Architecture, Engineering, and Construction (AEC) industry. Although still in its infancy, the notion of ``Construction 4.0'' such as DT can be driven by several technologies that are currently being applied across the different civil engineering phases (\cref{fig: civil engineering phases}). In this section, we provide a succinct review of these technologies and their limitations, delineating the need for a DT as a forward-thinking approach.

Qi et al.~\cite{Qi2021} describe construction as a production process, where the products are buildings. Associated with each product is a product-life cycle, which represents the conception till retirement of the product. In translation, the product life cycle of construction is conventionally categorized into the design phase, construction phase, operation and maintenance phase, and demolition phase. The design phase witnessed Building Information Modeling (BIM) as its major transforming technology and one of the building blocks for the new evolution. BIM signified a shift from traditional 2D computer-aided design (CAD) modeling to 3D parametric modeling with interoperability features (see \Cref{sec: 4}) for improved efficiency. Over the years, BIM has been universally adopted for the modeling and design of AEC projects. Doumbuoya et al.~\cite{Doumbouya2016} analyzed the utilization ratio of BIM in the industry and concluded that its rate of use is 55\% during the design phase, and 52\% during the detail design and tender stage. The authors also reported that the adoption of 3D modeling has resulted in a productivity gain of about 15\% of drawing production hours. Although BIM is ideally a life cycle process integration platform \cite{Arayici2009}, a major challenge in its adoption is its lack of sensitivity to change \cite{Doumbouya2016}. While numerous efforts have been employed to improve BIM functionality across different spheres \cite{Haekkinen2015,Wong2013}, there is a challenge with adopting it for already existing structures \cite{Khajavi2019}, i.e., as-built BIM model. In later sections, we will introduce the supporting mechanisms to drive BIM for such purposes and beyond -- all geared toward DT development and application. 

The construction phase is very vital as it is the phase in which the product comes into realization \cite{Opoku2021}. Just like every other phase, technological advancements have resulted in the need for automation to address some fundamental challenges that affect the timely, cost-effective, and efficient delivery of projects. Some areas of interest include logistics, scheduling, site monitoring, safety, and effective collaboration between stakeholders. For this phase, BIM with extensions is the major technological driver, but several limitations exist with this approach which are discussed in \Cref{subsec: 5.2}. Some of the driving technological adoptions later discussed include sensors and IoT (see \Cref{sec: 5}). There is, however, a major challenge to their integration for various construction management functions \cite{Opoku2021}, leaving a significant void in achieving the full automation of this phase. 

Structural health monitoring (SHM) is the dominant act of the operation and maintenance phase. Traditionally, this phase involved periodic visual inspections by field experts \cite{Gharehbaghi2022}. Currently, SHM approaches include computer vision and simulation-based techniques with sensors and IoT as major technological tools. Mishra et al.~\cite{Mishra2022} discussed cases of poor SHM resulting in disaster, necessitating the integration of sensing technologies such as wireless sensors, and IoT for improved monitoring processes and consequent disaster prevention \cite{Mishra2022}. Sun et al.~\cite{Sun2010} reviewed smart sensing technologies employed for SHM, and Ocheing et al \cite{Ochieng2018} assessed two ground-based radar technologies for a wind turbine. We once again hint that the realization of DT at this phase is more feasible and has witnessed greater application in comparison with the preceding phases. Finally, the demolition phase is rarely considered and has largely been neglected by researchers when considering the application of DT \cite{Opoku2021}. 

DT applications in several industries across product life cycles have exposed its potential in the civil engineering phases, with varying opinions on its method of adoption amongst researchers. Wagg et al.~\cite{Wagg2020} categorized their DT adoption more broadly into the design and asset management phase and discussed their proposed DT based on the product life cycle. Sacks et al.~\cite{Sacks2020} considered the workflow process of the DT. Others consider its employment in production, while some do not accede to any of these perspectives \cite{Negri2017}. Nonetheless, the relationship between DT with product life cycle and processes is commonly agreed upon \cite{Qi2021,Opoku2021,Wagg2020}. Ideally, the DT should span the entirety of the product life cycle from manufacture to retirement \cite{Wagg2020,Tuegel2011}. However, the analysis by  Liu et al.~\cite{Liu2021} showed that most DT papers focus on a single phase of the asset life cycle, with only 5\% addressing the full life cycle. In this paper, we will be adopting the three civil engineering phases as the project lifecycles and subsequently discussing their corresponding digital twin status and applications.

\section{Planning and Design: BIM and DT}
\label{sec: 4}
We commence this section by clarifying that the development of DT at the planning/design phase can be counter-intuitive due to the absence of a physical twin. Nonetheless, the concept of DT has promising applications for improved planning and efficient design of proposed projects within the AEC industry. It should be noted that while BIM is being applied across the three civil engineering phases, a practical outlook shows a comparatively heavy application in the design phase \cite{Doumbouya2016}. Moreover, BIM remains the ideal way to achieve a virtual representation of a proposed physical twin at the design phase, as opposed to a combined use of other technologies for DT development in the later phases. These perceptive realizations informed our segmented review of BIM and DT as they relate to the planning and design phase. In this section, we carefully selected specific applications of BIM that have evolved, all of which come together to enable the concept of DT using BIM at the planning and design phase. Importantly, reports on the techniques employed for BIM-based DT at this phase were presented, inspiring our views on the status quo and the way forward.

\subsection{Evolution of BIM}
\label{subsec: 4.1}
As a stepping stone to introduce the relationship between BIM and DT in the AEC industry, we begin by giving an evolutionary overview of BIM. To enable a robust overview, the evolutionary overview also implicitly reports the application of BIM for planning and design purposes at every stage of development. 

\subsubsection{Parametric Design}
\label{subsec: 4.1.1}
Prior to the development of BIM, traditional design approaches used CAD tools with pre-defined shapes for 2D and 3D modeling. For a wider range of applications and benefits, parametric design has been advanced by various researchers \cite{Katz2008,Venugopal2011,Halfawy2002} for design optimization. In contrast to the common-place modeling using pre-defined shapes, parametric design involves generating shapes from scratch through parametric input until the desired geometry is attained \cite{Girardet2021}. Parametric modeling allows for the extraction of multiple information---through a 360-degree visualization---of the model which increases information readiness for use in the analysis and optimization of designs. For example, Giradet et al.~\cite{Girardet2021} adopted a parametric BIM approach to enhance bridge project design and analysis. Their study developed a parametric algorithm to produce bridges on the Tekla Structures modeling platform. The advancements of BIM to include parametric design opened the doors to all-directional spatial modeling and BIM's advanced utilization for sustainable design.

\subsubsection{Sustainable Design}
\label{subsec: 4.1.2}
\begin{figure}[h]
    \centering
    \includegraphics[width=0.5\textwidth]{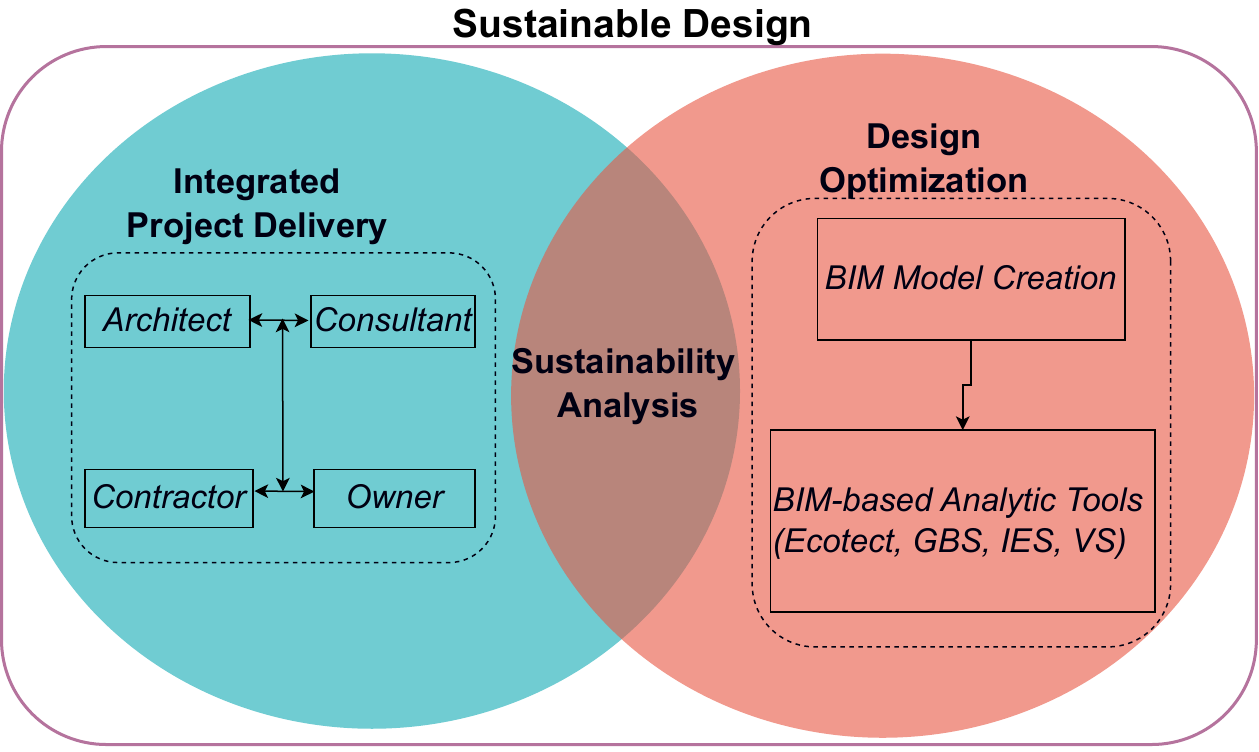}
    \caption{Sustainable design with BIM}
    \label{fig: Sustainable Design}
\end{figure}

Wong et al.~\cite{Wong2013} divided the use of BIM for sustainable design into two parts: Integrated Project Delivery and Design Optimization (\cref{fig: Sustainable Design}). The Integrated Project Delivery involves the use of BIM to foster easy communication among all practitioners in the AEC industry. Design optimization on the other hand is in two steps: the creation of models using inherent BIM software and the export of these models into BIM-based sustainability analysis tools. As typically done in engineering design, the results obtained from sustainability analysis are used for sustainable design. Hence, sustainability analysis---which covers various environmental analytical reports such as energy use, water use, daylighting, cost implications, etc.,---has become an important aspect of BIM tools.

Sustainability analysis implies the determination of a building's environmental performance using various tools for categorical investigation including building massing, site orientation, daylighting, and HVAC systems' maximization \cite{Azhar2009}. Autodesk Ecotect, Autodesk Green Building Studio (GBS), Integrated Environmental Solutions (IES), and Virtual Environment (VE) are the most commonly available tools. The application of BIM for sustainability analysis has increasingly become relevant over the years. This is due to rising pressures from environmental and social governance. As such, various countries have instituted the need for sustainability certification of major buildings through a credit-based system. In this regard, BIM can be used to acquire some credits from regulatory platforms such as the Leadership in Energy and Environmental Design (LEED), Building Research Establishment Environmental Assessment Methodology (BREEAM), and Building Environmental Assessment Method (BEAM), in the United States, United Kingdom, and China. In their research study, Wong et al.~\cite{Wong2014} reveal the important use of Autodesk Revit to generate significant credits for BEAM Plus (an accreditation platform in Hong Kong). Nguyen et al.~\cite{Nguyen2010} noted that BIM contains parameters that can be automatically acquired for sustainable design processes. A typical example is Revit's provision of material schedules that can be used for rating by LEED, BEAM Plus, BREEAM, and others.

\subsubsection{Interoperability and Semantic Models}
\label{subsec: 4.1.3}
Interoperability simply refers to the ability of BIM to export stored data into tools for further uses \cite{Azhar2009}. In the early days of BIM, several interoperability challenges necessitated standardization, which was successfully achieved by the body tagged Building SMART International through the development of Industry Foundation Classes (IFC) data models \cite{Lee2016}. IFC quickly became a success with the introduction of various data schemas such as EXPRESS schema for IFC-SPC file, XSD for XML file, and an alternative RDF for ifcOWL file (see \cite{Pauwels2015} for more details). Despite the capabilities offered by IFC classes through these schemas, several interoperability challenges persist across different application domains \cite{Lee2015}. Boje et al.~\cite{Boje2020} also noted that the IFC schemas were particularly created for data transfers from tool to tool with no allowance for dynamic modifications. Pauwels et al.~\cite{Pauwels2017} highlighted three other challenges associated with IFC schema which include binding, adaptability, and extensibility. Binding results from heterogeneous IFC translations from different authoring tools which lead to geometric distortion; adaptability is due to the not-so-flexible nature of the schemas for different application domains; and extensibility arises from the translation challenges experienced by third parties, who are not familiar with the EXPRESS language \cite{Pauwels2017}. 

In a bid to ensure dynamical collaborations of BIM models, the use of web-based semantic technologies such as \textit{Web Ontology Language (OWL)} models and \textit{Linked Data (LD)} were investigated and subsequently developed \cite{Boje2020,Pauwels2017,Le2016, Venugopal2015,Pauwels2016}. For example, IfcOWL was introduced as alternative support for BIM interoperability using a web-based semantic model, which Pauwels et al., \cite{Pauwels2016} proposed a recommendable version for commercial use. OWL and LD have received positive responses from researchers and industry experts in the AEC industry as it is well inclined to the pervasive nature of big data in the Internet of Things (IoT) era. Boje et al.~\cite{Boje2020} argued that adopting OWL, LD, and in general semantic web worldview is a requisite for the continued relevance and improved value creation of BIM to the AEC industry. It should be noted that the concise overview of semantic models from OWL and LD perspectives is intended to familiarize readers with an important aspect of data transmissibility in DT via BIM and sensors (more details will be provided in \Cref{subsec: 5.2}).

\subsubsection{Static to Dynamic: BIM and DT}
\label{subsec: 4.1.4}
\begin{figure}[t]
    \centering
    \includegraphics[width=0.5\textwidth]{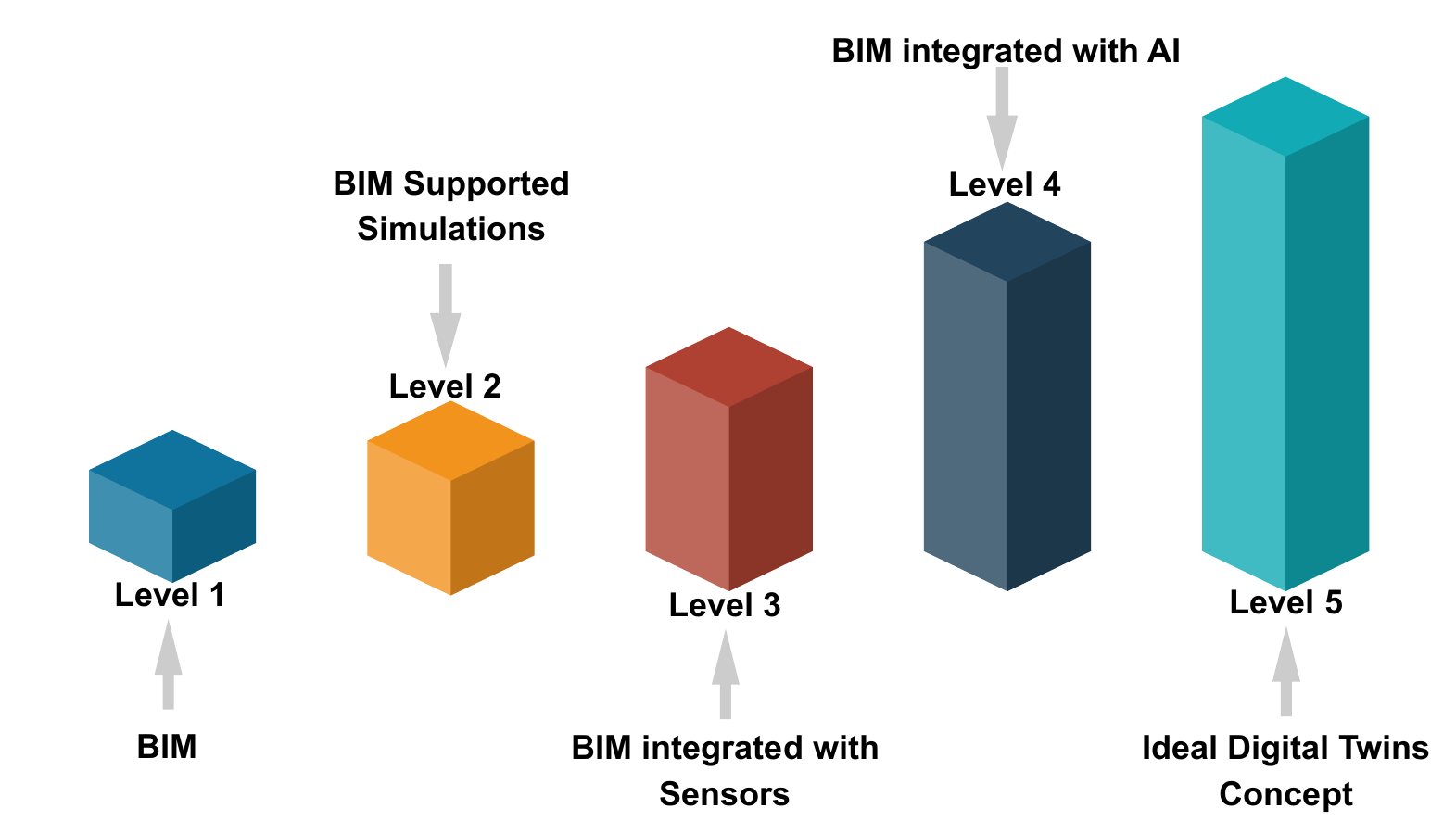}
    \caption{Static to dynamic capabilities of BIM (based on \cite{Deng2021})}
    \label{fig: Static_to_Dynamic}
\end{figure}

We utilize the five-level framework proposed by \cite{Deng2021} to delineate the evolution of BIM from its inherent static configuration to an integrated dynamic powerhouse as an enabler for DT development in the built environment. The five levels as shown in Fig.~\ref{fig: Static_to_Dynamic} involve the following: BIM general applicability, BIM plus simulation, BIM plus sensors, BIM plus artificial intelligence (AI), and DT. A worthy observation is that these levels, although not explicitly stated by \cite{Deng2021}, also depict the chronological advancement of BIM integration for DT.

Level 1, termed BIM general applicability, involves the advancement of traditional design mechanisms by enabling better visual representations and information sharing across various building life cycles. However, the applications at this level only utilized static information provided by BIM, limiting their use for design integration and consequently motivating Level 2, termed BIM-supported simulations. This advances the general use cases from level 1 to include simulations and provides an analytic overview of building performances for optimal design insights \cite{Deng2021}. We notice from available literature \cite{Zhang2016,Lee2016,Tagliabue2018,Shahzad2019,Ma2019,Cheng2014,Wu2013} that BIM-supported simulations give rise to sustainability analysis as their application revolves around energy performance evaluation, green building design, day-lighting analysis, and acoustic simulations as highlighted in \Cref{subsec: 4.1.2}. The bottleneck of these diverse use cases reported for level 2 is their sole dependence on static information with no regard for potential changes (i.e., in real time) at the later part of the structures.

\begin{figure}[t]
    \centering
    \includegraphics[width=0.3\textwidth]{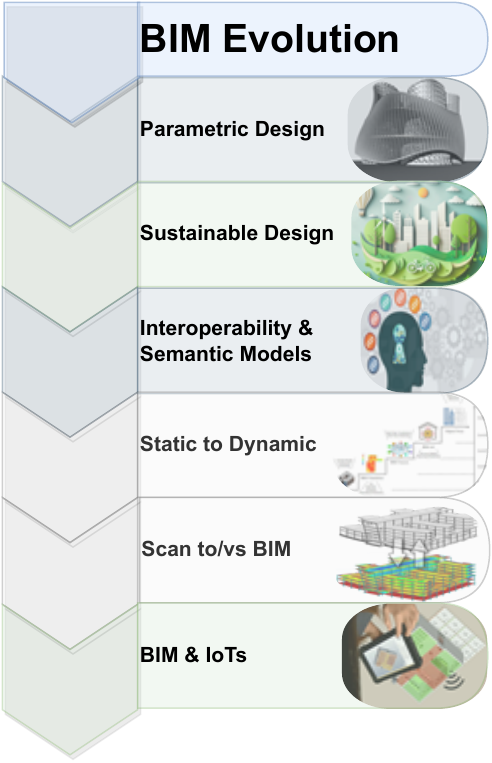}
    \caption{Evolution of BIM}
    \label{fig: BIM Evolution}
\end{figure}

Level 3, termed integration of BIM and IoT techniques, became a gamechanger in enabling real-time monitoring of building performances, thanks to technological drivers such as sensors, IoTs, data analytics query, semantic models, LD, OWL, relational database, and global/universal unique identifiers \cite{Rasheed2020}. They utilized time-series data from sensors and contextual data or semantics from BIM (as detailed in \cref{subsec: 5.3} \cite{Tang2019}).  Level 4, termed BIM and artificial intelligence (AI) integration, takes level 3 a step further to facilitate easy decision-making. Deng et al.~\cite{Deng2021} termed level 5 as a conceptual framework of an ideal DT in the built environment. The major takeaway from \cite{Deng2021}'s review is the need to enable automated feedback control beyond level 4 which stops at real-time predictions but still needs humans in the loops for primitive decision-making. While the review provided by \cite{Deng2021} was termed BIM evolution, we believe that such characterization is limited in revealing the many other groundbreaking applications of BIM towards DT development. Beyond the dynamical application of BIM, some advances concerning monitoring, prediction, and control have become a field of their own. The relevant ones (as shown in \cref{fig: BIM Evolution}) are further presented as the remaining sub-sections of the evolution of BIM.

\subsubsection{Scan-to-BIM and Scan-vs-BIM}
\label{subsec: 4.1.5}
Noting that one of the major edges of BIM is the ability to produce semantically rich representations of models, efforts have been put in place to produce BIM models for not only proposed projects but also completed projects which are respectively called as-designed BIM models and as-built BIM models. \textit{As-designed BIM model} simply refers to the creation of BIM models from design conditions while the \textit{as-built BIM model} (formally called as-built/as-is) refers to the creation of a BIM model from an existing facility. The as-built model involves acquiring geometrical information from the built structure and their consequent transformation to a high-fidelity model with semantically rich representations \cite{Tang2010}.

The creation of as-designed BIM models has become primitive as various design conditions can be richly represented on BIM software utilizing the standardization from IFCs, i.e., interoperability. On the other hand, the creation of as-built BIM models requires significant human expertise and requisite technological tools - one of which is 3D Laser Scanning (also called LADAR meaning laser detection and ranging). LADAR generates a 3D point cloud of a visible scene which can then be processed for further uses like monitoring, geometrical modeling, and in turn developing an as-built BIM model. The process of creating 3D BIM models from acquired 3D-point clouds is regarded as ``Scan-to-BIM'' \cite{Bosche2014}. For such a technique, laser scanning has been well adopted in the AEC field as it is arguably considered to be the best available technology for capturing 3D information on a project with high accuracy and speed \cite{Bosche2014}.

Tang et al.~\cite{Tang2010} rigorously reviewed Scan-to-BIM processes with a focus on automating the difficult manual processes. From the authors' review, the creation of an as-built BIM using laser scanners involves three core steps which include data collection, data preprocessing, and modeling. The data collection is done by laser scanning technology which measures distances from sensors and nearby surfaces to give 3D point clouds. These point clouds which could be in several millions are then aligned from the scanner's local coordinate system to the global coordinate system (a process known as registration). Filtering can also be carried out in the data preprocessing stage. The modeling state then involves geometric modeling, object categorization, and relationship modeling.

One would notice that the process of Scan-to-BIM does not involve the utilization of prior information from the as-designed models to develop the as-built model. Tang et al.~\cite{Tang2019} pointed out that one of the ways of automating object recognition is using prior knowledge to reduce the search space of object recognition algorithms. In a bid to achieve such a level of automation coupled with the rapid development of 3D BIM modeling, Bosche et al.~\cite{Bosche2014} developed the ``Scan-versus-BIM'' approach. This involves the alignment of laser scans of construction sites with as-design 3D BIM models, that then use comparative analysis based on proximity metrics to automatically recognize and uniquely identify 3D model objects. Unlike Scan-to-BIM which recognizes geometrically similar objects as one object, Scan-versus-BIM uniquely identifies these objects as separate objects in the 3D BIM model, hence the edge of the latter over the former \cite{Bosche2014}. Readers can refer to Bosche's review on the four-step process of Scan-versus-BIM for unique object recognition. The Scan-versus-BIM approach is however limited by their specific use for objects within proximity which is usually set at a maximum threshold ($\delta_{\max} = 50 \mathrm{mm}$). This prevents a flexible application of the method. In a later work, Bosche et al.~\cite{Bosche2015} integrated both Scan-to-BIM and Scan-versus-BIM approach for construction monitoring using cylindrical MEP components as a case study.

\subsubsection{BIM and IoTs}
The integration of BIM and IoT are complementary as the former gives a high-fidelity representation of the as-designed buildings giving rise to geometrical and semantic information. IoT takes such developments further by giving a status report by providing real-time information during the subsequent life cycles of buildings, especially during construction and operations \cite{Tang2019}. We hint that IoTs significantly advanced the construction monitoring process. To these foregoing, a detailed overview of BIM and IoTs including the concept of relational and contextual databases will be presented in \Cref{subsec: 5.2} as it relates to BIM integration and subsequent DT development.

\subsection{Techniques for DT Adoption}
\label{subsec: 4.2}
Based on the various applications of BIM previously presented, we now establish the conceptual and practical realization of DT using BIM. We particularly presented a dual review by first noting the remarkable use of BIM for DT creation and subsequently establishing our inference on the status quo of BIM and DT for the planning and design phase in the second part.

\subsubsection{BIM-based DT Developments and Applications}

Lu et al.~\cite{Lu2020a} developed and applied a framework hinged on four practical requirements (intelligence, efficiency, integration, and interoperability) to develop and utilize DT for smart asset management. In their proposed framework, three key layers were presented including the smart asset layer, smart asset integration layer, and smart DT-enabled asset management layer. In these layers, sensing, IoT, as-is BIM models, cloud computing, and computer vision are important technologies. In a separate work, Lu et al.~\cite{Lu2020} extended their framework development to the development of a DT at the building and city levels by presenting a hierarchical architecture. The architecture comprised five layers: data acquisition layer, transmission layer, digital modeling layer, data-model integration layer, and service layer. DT at city levels was imagined as a parent umbrella to other sub-DTs from various sub-assets that make up the city of interest. BIM was particularly employed in the digital modeling layer in combination with City Information Modeling and supplementary information such as weather and historical data. The authors demonstrated their architecture by developing a dynamic DT for the west Cambridge site of the University of Cambridge (see \cite{Lu2020} for more details). In the same trend, further works were carried out for DT-enabled anomaly detection for built asset monitoring in operation and maintenance (see \cite{Lu2020b}).

In another spectrum, Pan et al.~\cite{Pan2021} developed a BIM-data mining integrated DT framework for advanced project/construction management. Due to IFC interoperability limitations, extensible techniques such as retrieval of CSV files (called event logs) from IFC files are adopted for subsequent data mining, otherwise tagged process mining. This technique was combined with BIM and IoT in pairing the virtual and physical models for advanced construction management. They presented an architecture of the proposed DT which comprises: sensing technologies for data collection from the as-built environment, cloud storage for information storage and a bridge of the physical-cyber system, and high fidelity models including 4D visualization and process model for modeling, bottleneck detection, and construction progress prediction. Within the BIM cloud, the point cloud data from the sensor is compared with the as-planned IFC using a tool named Real-Time and Automated Monitoring and Control (RAAMAC) to establish discrepancies between as-planned and as-built models. The data from this tool is typically in IFC format which is unreadable by data mining algorithms and as such an IFC logger is used to parse required data from IFC to event logs in CSV format. With such events log in place, model construction, model implementation, and model evaluation were carried out. Pan et al.~\cite{Pan2021} then presented a case study of a BIM-based construction project for analysis, prediction, and optimization.

In a bid to balance the parallel developments of BIM from the AEC industry and DT from the manufacturing industry, Badenko et al.~\cite{Badenko2021} present the integration of DT and BIM technologies within the factories of the future (FoF). FoF uses systems of complex technological solutions that provide the shortest possible time for the design and production of quality products. The authors highlighted three principles as a basis for the integration of BIM and DT technologies for FoF. The first principle emphasizes flexible production technologies and infrastructure for dynamic product development. The second notes the validation of smart models via virtual testing, sensing technologies, and Digital Shadow (DS) principles. The third implies using three key technologies including BIM, DT, and System Information Modeling (SIM) for increased economic benefits. SIM simply involves the transformation of management systems for products, technologies, and production infrastructure as objects of management. On these principles are the integration methods for DT and BIM for FoF. They involve the construction of a matrix of target indicators and resource constraints of the entire FoF and an information management system towards the creation of an accurate and reliable model. The creation of such a model must be coordinated via the increasing levels of virtual models across graphical data (BIM), non-graphical data, and documentation, all geared towards the creation of an Asset Management Model.

\subsubsection{BIM-DT for Design Optimization}
The aforementioned applications of BIM-based DT focused more on the construction, operations, and maintenance phases of AEC projects. While BIM has successfully been applied in the planning/design phase towards sustainable design as earlier reported, DT explicit use in the same regard has been limited in literature. The major setback of BIM-based DT for planning and design can be linked to the absence of a real physical twin at this phase, thus making the idea of a DT not fully realizable. Nonetheless, we crucially report that DT can also aid in design optimization. For this, the creation of a temporal DT at the design phase by using similar or historical structures as corresponding physical twins is being advanced \cite{Jiang2021}. The existence of such DT will enable proper planning and insightful designs based on the visualized data from the existing DT. For example, to achieve the optimized design of a bridge, a DT of an existing bridge with similarities such as location, functional, and aesthetic requirements can be developed as a starting point for the proposed bridge structure. Such application of DT will become particularly important for high-risk projects or conditions as a means to reduce uncertainty while optimizing cost reductions.

\subsection{Remarks on BIM-based DT for Civil Engineering Phases}
\label{subsec: 4.3}
The creation of DT at the planning/design phase is rather conceptual than practical, but, such development has been justified for its use in design optimization as previously presented. However, the possibilities offered by BIM---as reported in \Cref{subsec: 4.2} pave the way for DT realizations in subsequent phases of civil engineering projects. Notable mentions include the semantic richness of BIM to enable comparative monitoring (see \Cref{subsec: 5.4}), Scan-to-BIM and Scan-vs-BIM for robust visualization and thus enabling DT creation for legacy building (see \Cref{subsec: 6.6}), dynamic BIM for real-time monitoring and predictions, among others. To these foregoing, we introduce three classifications of DT with respect to their use at different phases of civil engineering projects comprising: indirect DT, partial DT, and fully realized DT. The term \textit{indirect DT} implies the use of similar or historical structures as a physical twin to develop a temporal DT of a proposed project at the design phase. \textit{Partial or Prototype DT} refers to transient DT at the construction stage due to a high level of variability in the physical twin. In this case, BIM, which is often far ahead of the physical twin, is often used as a digital model representation in addition to enabling feedback and control to realize DT. Fully realized DT is apparent at the O\&M phase as efforts are dedicated to realizing the operational conditions of the physical twin in DT as opposed to both building the physical twin and updating the DT in the other two phases. Sections \ref{sec: 5} and \ref{sec: 6} respectively present DT development at the construction, and O\&M phase. 

Taking cues from \cite{Badenko2021}, we further present the evolutionary developments of BIM and their respective relation to DT on a scale basis. Let us be clear, we do not attempt to present BIM as DT but rather reveal that a fully-realized DT should perform the functionalities of a BIM and beyond. While BIM gives full semantics and 3D/4D visualization, DT distinguishing features comprise two-way coupled connections between the physical twin (existing structures) and virtual twin, thereby enabling feedback and control as hinted in the level 5 of static to dynamic evolutions in \Cref{subsec: 4.1.4}.  In relation to our adopted DT classification on the digital model, digital shadow, and digital twin, a parallel scale for BIM from level 0 to 3 was introduced by \cite{Jones2020} and presented by \cite{Badenko2021}. In this regard, BIM-level 0 comprising only 2D CAD drawings does not come into the DT levels at all. Subsequently, the BIM-level 1 which consists of 3D models comes as a Digital Model in the DT scale. A combination of 3D and real-to-virtual connection was termed BIM-level 2 and Digital Shadow respectively and a combined 3D model with bi-directional real-to-virtual and virtual-to-real connection was staged BIM-level 3 and DT. 

\section{Construction: Sensing, IoT and DT}
\label{sec: 5}
The technological evolution of DT for construction process optimization cannot be made possible without the tremendous developments in construction monitoring over the years. Although DT aims to advance the capabilities of construction monitoring processes, the fundamental techniques employed in the latter such as the use of sensing technologies and IoTs are important recipes for the former realizations. To establish the development and application of DT in the construction phase, it is crucial to avail expositions on these technological drivers. While these technologies are applied in the subsequent phase, the full commencement of their application can be initially linked to the construction phase, hence, we present them in this section. Also, we report, in the later part of this section, developed or applied techniques that reflect level 3--4 DT capabilities, and as stepping stones to realizing the actuating capabilities of DT for the construction phase.

\subsection{Sensing}
\label{sec: 5.1}
Sensing technologies paved the way for real-time construction monitoring offering diverse advantages for construction site management such as improved productivity, hazard identification \cite{Zhao2021}, construction waste management \cite{Sartipi2020}, clash detection, discrepancies detection \cite{Ahmed2020} among others. Rao et al.~\cite{Rao2022} already provided a detailed and comprehensive overview of sensing technologies. The three-phased review done by \cite{Rao2022} includes sensing construction environment, real-time monitoring methodologies, and case studies of construction site monitoring. The sensing part of their review is divided into three aspects including mapping sensors, positional and communication sensors, and sensor platforms. Following this classification, we summarize notable sensors in the following paragraphs that are mostly employed in construction and would be useful for DT development. But first, we established some basic definitions in Table \ref{sensor-terminologies}.

\begin{table*}[t]
    \caption{Basic terminologies used in sensing technologies.\label{sensor-terminologies}}
    \centering{%
    \begin{tabular}{  p{2.8cm}  p{13.4cm} }
    \toprule
    \textbf{Terminologies} & \textbf{Definitions} \\ 
    \midrule
    3D point cloud & a set of data points defined in a 3D coordinates system i.e., by their X, Y, and Z coordinates. A point cloud delineates a single spatial measurement of an object's surface, and many points represent the entire surface of the target object. Color intensities can also be added to a point cloud. They are usually created by sensing technologies such as 3D laser scanners or photogrammetry software \cite{ElOmari2011,Gigabyte2024}. \\ [1.3cm]
    Photogrammetry & This involves extracting information from an object's measured photographs (mainly establishing the geometric relationship between image and object at a specific time) \cite{Mikhail2001}. \\ [0.5cm]
    Mapping sensors & involves data capturing (3D point cloud) using sensing technologies that employ physics-based techniques such as electromagnetic radiation and images combined with geometric estimation (such as Time-of-Flight), and modeling techniques (such as photogrammetry) to capture the as-is condition of a facility or construction site \cite{Rao2022}. \\ [1.3cm]
    
    Positional and Communication Sensors & They are used to obtain the geographical locations and motion of the sensors, which are used to georeference the collected 3D data through joint use of measurements from different sensors otherwise called fusion \cite{Elhashash2022}. \\ [1cm]
    
    Sensor platforms & The means of structural supports provided for sensors. They can be stationary or mobile. \\
    \bottomrule
    \end{tabular}
    }%
   
\end{table*}

For mapping sensors, we highlight three commonly used types which comprise laser scanners, RGB cameras, and depth cameras. Laser scanners utilize a laser to generate a visible spectrum for calculating the emitted pulse Time of Flight (ToF). Laser scanners are however limited by the lack of color information, hence their combination with digital cameras to obtain colorized 3D point clouds. Regardless, laser scanner (specifically terrestrial laser scanners or TLS) is often regarded as the best mapping sensor type in the construction industry due to their high accuracy \cite{Rao2022}. RGB cameras use photogrammetric techniques to process digital images and thus create 3D point clouds. However, RGB accuracy depends on several factors such as camera model, camera resolution, scene illumination, and focal length, among others \cite{Dai2013}. As such, their accuracy is lower than 3D laser scanners. Nonetheless, the adopted photogrammetry technique is more flexible and efficient which helps retain RGB cameras' competitiveness in the construction industry. Depth cameras are specific types of cameras equipped with range sensors such as ToF sensors to illuminate the whole scene with one pulse of radiation and capture all the reflected lights at once. The range data obtained are then converted to depth data to represent distances which are further augmented to RGB-D images containing color and depth information for each pixel. Depth cameras are limited by a small range of accuracy (3 to 10m) and are such are only suitable for small indoor environments. However, their affordability and miniature size make them useful in mobile devices \cite{Rao2022}. The aforementioned sensors can be used for tracking construction sites, safety management, indoor localization, 3D point cloud generation, 3D modeling, and DT development. 

Positioning and communication sensors include identification and tracking devices such as Barcodes and Radio Frequency Identification (RFID), inertial measurement units (IMU) consisting of accelerometers, gyroscopes, and magnetometers for 3D measurements, and global navigation satellite systems (GNSS) such as the commonly adopted United States Global Positioning System (GPS). Other positioning and communication sensors include short-range communication technologies such as Wi-Fi and infrared (IR) sensing, and long-range communication technologies which comprise Frequency Modulation (FM) and cellular communications (see \cite{Rao2022} for more details). Also, sensor platforms could be mobile or stationary. Some notable mentions include handheld, wearable, trolley, and Unmanned Ground Vehicle (UGV), all of which can be either used separately or combined for specific use cases \cite{Price2021}. It should be noted that the choice of platforms greatly influences the applications and performances of the sensors, necessitating a reconnaissance survey of the sites for appropriate sensor and platform selections.

\subsection{Internet of Things (IoT)}
\label{subsec: 5.2}
Numerous research has advanced the role of IoT in the construction industry for automated processes. The same trend can also be seen in review articles \cite{Woodhead2018,Khurshid2023,Tang2019,Ghosh2021,Oke2021} that not only advanced the revolutionary roles of IoT for automated construction processes but also reported several case studies that have utilized IoT for the same. In a trend analysis study by Ghosh et al.~\cite{Ghosh2021}, four clusters including structural health monitoring, construction safety, optimization and simulation, and image processing were identified as prominent areas of research that apply IoT in the construction industry. We will not explore such benefits as readers can always refer to cited literature but rather show a narrative that reveals the possibilities of DT for construction process optimization.

The use of IoT in the construction phase is not commonplace when compared with the O\&M phase or even some part of the design phase. This is mainly due to the complexities associated with the fast-paced nature of the construction phase. A major complex scenario is data heterogeneity challenges from sensors and virtual models (i.e., BIM). More concretely, sensors generate time series data from continuous sensor readings usually stored in a relational database \cite{Kazmi2014}, while the BIM model generates contextual data in IFC formats, presenting a challenge of data integration let alone enabling dynamical updates via the Internet. To these foregoing, we present important details from the available review by Tang et al.~\cite{Tang2019} on BIM and IoT integration evolutionary methods over the years. But first, it is important to clarify some important terminologies including a self-contained definition of IoT. 

\begin{itemize}
    \item What is IoT? Tang et al.~\cite{Tang2019} identified that the key concern of IoT is the interconnection of sensing and actuation devices that enable information sharing via the Internet. In a similar wave of definitions, Gubbi et al.~\cite{Gubbi2013} defined IoT as an interconnected network of physical objects with sensing, actuating, and communication capabilities that enable a central framework for data syntheses via seamless access to domain-specific software and services. While these definitions connote the concept of IoT, the explanation given by Woodhead et al.~\cite{Woodhead2018} is perhaps the simplest and most relatable. They implicitly assume that the word ``Internet'' is commonplace but elucidate more on the corresponding word ``Thing'' that makes up ``Internet of Things''. As presented, a ``Thing'' is simply any object that has an Internet Protocol (IP) address and sends data about its state and immediate environment, and can also receive data linked to actions. They further presented the emergence of IoT as a means to converge the developments associated with information technology and operational technology which opens the way for embedding compute power in or on ``things'' such as sensors. 

    \item Industrial IoT (IIoT): This refers to the extended use of IoT technology in industrial settings through cloud services and virtualization. This has become an important enabler for real-time representations and control of complex systems towards digital constructions \cite{Oracle2023,Woodhead2018}.

    \item Ubiquitous Computing/Connectivity: This implies the all-round integration of computation capabilities into the physical environment, advancing the perception of the physical environment beyond just a visible object or representation of a model. This is made possible by placing sensor networks everywhere to perform an ad hoc arrangement that offers a higher degree of flexibility for settings and an increased number of sensor nodes \cite{Skibniewski2006}. 

    \item Contextual and Relational Database: The former as the name implies gives context to an entity, person, or thing. In the AEC industry, the BIM is the main repository for contextual data which provides information about building geometry, IoT devices' configuration, occupancy information, and weather forecast among others. Precisely, the contextual data from BIM are stored in a database typically in IFC formats. On the other spectrum in the construction sites, time series data which records continuous sensor readings are stored in a relational database \cite{Tang2019}. A relational database stores related data points and provides fast access to them. They are typically organized into tables, and each row is a record with a unique ID (key) and the corresponding columns depict the attributes of the data. Relational databases are built to understand Structured Query Language (SQL). 

    \item Database Schema: a collection of metadata that describes the structures and organization of a relational database including constraints such as tables, names, fields, data types, and their respective relationships \cite{IBM2023}.
\end{itemize}

\subsection{BIM and IoT Devices Integration Methods for Construction Process Monitoring}
\label{subsec: 5.3}

The integration methods between BIM contextual data and sensor time-series data (otherwise tagged relational data) for construction site monitoring is a critical aspect of DT development for construction process optimization. Three out of the five integration methods highlighted by Tang et al.~\cite{Tang2019} have demonstrated successful integration of BIM and IoT \cite{Zhang2015,Solihin2017,Mazairac2013}, but not necessarily on a large scale or in real time. These methods include the combination of BIM tools' Application Programming Interface (API)---such as Revit DB Link---and relational database---such as Microsoft Access; the transformation of BIM contextual data into a relational database using new data schema; and the third is the creation of a new query language. The remaining two are distinct as they enable the realization of IoT concepts due to their ability to process heterogeneous data through a unified framework via the Internet. One of such is the semantic web approach which processes heterogeneous data---i.e., both contextual and time series data---into a homogeneous data tagged Resource Description Format (RDF). The different data silos are then linked across different domains via unique identification which are followed by data queries that can be presented in different forms. However, the conversion of all data types into RDF takes away some inherent advantages of databases in their original forms. In particular, RDF data structure is less effective when compared with relational database structure, and data duplication during data conversion to RDF is a probable challenge \cite{Hu2016, Tang2019}. To circumvent such setbacks, the fifth and most robust BIM-IoT integration method evolved which simply merges semantic web and relational database.

\begin{figure}[t]
    \centering
    \includegraphics[width=0.5\textwidth]{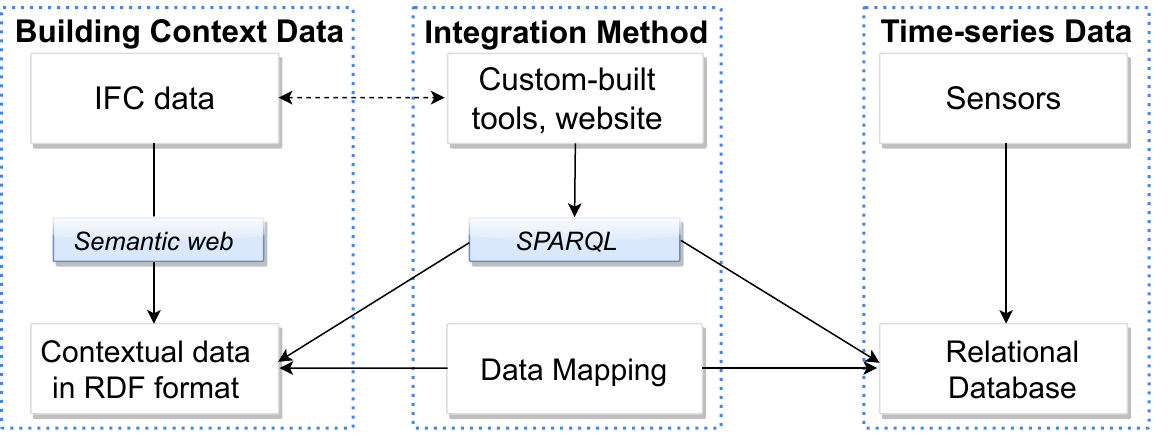}
    \caption{Semantic + Relational Database BIM-IoT integration method (Adapted from \cite{Tang2019}).}
    \label{fig:Semantic+Relational}
\end{figure}

The semantic web plus relational database integration method retains both contextual and relational databases in their original format. The contextual data does not only contain building-related information but also sensor information such as sensor tags and other semantic representations (but not sensor readings). The process involves representing the contextual information in RDF format using a semantic web approach, retaining sensor time-series data in the relational database, and mapping contextual information with the time-series data which can be referenced using Sensor ID described in RDF as part of the sensor semantics in the contextual database. The combination of two technologies necessitates an integrated query method in which the contextual data are queried by SPARQL while the time series are queried using SQL. Since both data types are mapped, SQL queries can be created using SPARQL queries on RDF data (see Fig.~\ref{fig:Semantic+Relational}). Beyond the obvious advantage associated with retaining data in their original form, the method is time-saving and memory-efficient. Also, the integrated query method makes use of trusted query languages which include SPARQL and SQL. As such Tang et al.~\cite{Tang2019} described this approach as one of the most promising methods to facilitate IoT deployment in the construction industry.

\subsection{Techniques for DT Adoption}
\label{subsec: 5.4}
 DT in construction has advanced mere reporting capabilities to enable automated and intelligent semantic platforms. Some current adaptations of the individual components of the DT for construction process enhancement are presented in this subsection. 

 \subsubsection{Automated Site Progress Monitoring}
Globally, automated construction progress monitoring consists of data collection, comparison of as-planned and as-built models (Scan-vs-BIM) for progress evaluation (see \Cref{subsec: 4.1.5}), and visualization of results. As the starting point of this application, the devices used for data collection (typically sensors) significantly influence the success of automated site progress monitoring \cite{GolparvarFard2009,Alaloul2021}. The progress evaluation component can be achieved using laser scanning and photogrammetry to generate point clouds, allowing temporal elements not reflected on the BIM to be captured \cite{Braun2015}. The last element of automated construction monitoring is visualization. Its driving technological advancement is Augmented reality \cite{Kopsida2015,Alaloul2021}. By \textit{Augmented Reality}, we refer to an environment where users can view virtual elements in the real world \cite{Alaloul2021}. The integration of augmented reality with BIM allows for swift site updating of 4D BIM models using smartphones. It also permits the superposition of the as-built models on the as-planned models. However, Alaloul et al.~\cite{Alaloul2021} highlighted the inapplicability of this method for remote monitoring as a deficiency.

Braun et al.~\cite{Braun2015} carried out automatic progress monitoring of a 5-story office building using photogrammetric surveys and 4D BIM. Point clouds were created from the surveys using semi-global-matching and subsequently coordinated with the 4D BIM model. Pucko et al \cite{Pucko2018} proposed a continuous construction progress monitoring method, which is sensitive to change and constantly updated during the construction process. They utilized small 3D scanning devices placed in the helmets of workers for real-time data capturing and recording of partial point clouds. Furthermore, they attempted to provide a holistic progress monitoring method that caters to both interior and exterior progress monitoring. Interior automated progress monitoring has been identified as a particularly challenging endeavor for a myriad of reasons such as disparity in the appearance of objects, lack of texture of some indoor elements, lighting issues, presence of transparent elements and sizing of indoor elements, and difficulty in identifying the position of elements, all of which hinder the extraction of features by data collection devices \cite{Kopsida2015,Ekanayake2021}. Ekanayake et al.~\cite{Ekanayake2021} conducted a detailed analysis of computer vision-based interior construction progress monitoring. The trend analysis executed by the authors corroborates the gap between exterior and interior progress monitoring in construction. Similarly, Reja et al \cite{Reja2022} carried out an in-depth analysis of current adoptions of automated progress monitoring technologies in construction, and highlighted the pros and cons of each technology such as optimal location tracking and difficulty in indoor monitoring of geospatial devices such as GPS respectively. The authors identified DT as a potential solution for addressing some of the challenges.

In general, the major challenges plaguing automated construction progress monitoring include fragmentation between interior and exterior progress monitoring, fragmentation of monitoring process, lack of integration of input data, occlusions, and cost. Despite the contributions of researchers to digitalize the monitoring process, most methods available still require some manual input in the completion of the monitoring process. Based on the DT classification adopted in this paper, the DT in construction progress monitoring is between levels 2 and 3. Thus, a construction DT with a complete loop (i.e., level 5) that is capable of prediction, interactions with the environment, and is highly autonomous or requires limited human inputs will sufficiently tackle the challenges identified above.

\begin{figure}[h]
    \centering
    \includegraphics[width=0.3\textwidth]{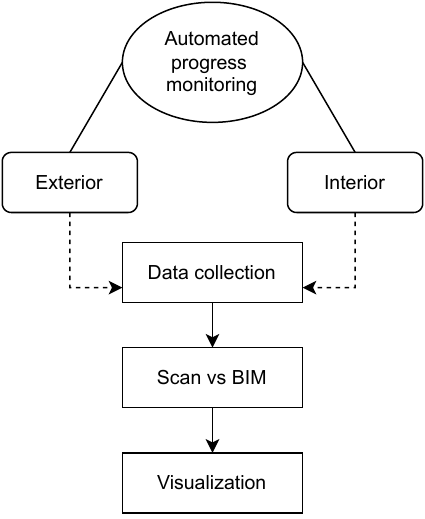}
    \caption{Elements of automated construction progress monitoring and corresponding DT scale. \textnormal {\textit{Bold arrows represent a sequential ordering. Dashed arrows represent a procedural relationship. Straight lines indicate categorization.}}}
    \label{fig:DT + Automated construction progress monitoring}
\end{figure}

\subsubsection{Hazard Identification and Construction Safety}

The conceptual use of digitization for hazard identification and construction safety is prominent among researchers. These techniques although not explicitly stated have close relationships with the proposed use of DT for similar purposes. Hou et al.~\cite{Hou2020} in a review on DT applications for construction workforce safety reported the existing approaches such that they fit into the accident causality framework which includes construction environment safety, behavior safety in construction, and safety awareness in construction. The authors identified sensor technologies, virtual construction simulation (VCS), and visualization technology as the enabling technologies for construction workforce safety. They further sub-categorized the use of sensor technology as an applied DT concept for construction workforce safety into hazard identification, on-site individual positioning, and workforce behavior monitoring, all of which computer vision was reported as the most popular technological method for extracting hidden information from complex data. Soltani et al.~\cite{Soltani2018} utilized several construction site images to estimate the 3D pose of an excavator. Park et al.~\cite{Park2016} and Konstantinou et al.\cite{Konstantinou2018} separately utilized computer vision algorithms to identify and track individuals and objects. Also, Park et al.~\cite{Park2015} utilized motion monitoring to identify the position and hard hats usage by workers on the construction site. On another spectrum, the use of VCS and visualization technology (such as BIM) was identified to help in safety planning, construction activity visualization, and safety training. For example, some researchers utilized logic-based mapping of safety rules in BIM for automated hazard identification and countermeasures \cite{Zhang2013,Getuli2017}. 

A closer observation indicates that the majority of the highlighted benefits do not complete the loop for automated feedback and control which remains an unattained vision of DT for the construction industry in general. However, we acknowledge that the several demonstrations already reviewed by Hou et al.~\cite{Hou2020} attempt to go beyond hazard identification by demonstrating safety awareness and response options available to construction workers. However, their complete realizations were limited by the lack of robust methods for complex data gathering and synchronization, inefficient information processing of complex logical relationships among objects, hazards, and safety rules, and unclear warning mechanisms that best align with the cognitive abilities of workers for proper safety behaviors and response. Some recommended techniques to address these challenges include simultaneous localization and mapping (SLAM), ontological-based relationship modeling, and robust warning mechanisms for improved workers' cognitive safety activities \cite{Hou2020}. We also recall the use of robust BIM-IoT integration methods recommended earlier presented in \cref{subsec: 5.3} to tackle some of the highlighted challenges.

Recently Ye et al.~\cite{Ye2023} developed a DT-based multi-information intelligent early warning and safety management platform for tunnel construction. The physical twin was equipped with several sensors, a gateway, and Wireless Sensor Networks (WSNs) transmission lines to achieve multiple information-gathering and real-time connections between construction workers and managers during tunnel construction. The DT was developed on the physical layer through internet connectivity for real virtual construction simulation, which was further augmented with a service layer for managerial access and control. In addition, a four-level safety early warning and corresponding emergency response plans were provided for improved decision-making. The model was tested on a real-life project in China, to successfully predict the collapse of a tunnel excavation and the proactive rescue of construction workers. Nonetheless, the developed DT is limited in terms of robust information processing for safety rules and the inability to avoid accidents rather than predicting the same \cite{Ye2023}. We also believe that automated safety measures update in the model must be considered in future research and continued reliance on manual rule checking and updates could limit broader application of the proposed technique.

\subsubsection{Clash Detection and Simulation}
In construction, clashes may refer to geometric incompatibilities between corresponding design contributions of various AEC stakeholders, which if left unchecked would result in cost over-run, waste of resources, and time delays \cite{RachaChahrour2021,Seo2012}. These geometric inconsistencies mainly occur due to the overlay of plans from multi-disciplines \cite{Rokooei2015}. Some other attributed causes of clashes comprise design uncertainties, design complexity, model tolerance, use of 2D models, incorrect detail, isolated designs by stakeholders, and design errors \cite{RachaChahrour2021,Akponeware2017}. To circumvent the consequences of clashes in construction projects, the concept of clash detection emerged and it was traditionally conducted by manually comparing the 2D drawings of all stakeholders to determine areas with collisions or clashes \cite{Hartmann2010,Hu2020,Helm2010}. As a result of the emergence of BIM, the automation of clash detection has been enabled. A \textit{Federated Model} can be created via BIM, which is an integrated model of all the multi-disciplinary 3D models. The geometric information from the federated model can be employed to compute regions with clashes \cite{Hu2020,Gholizadeh2018}. 

Researchers have invested efforts into improving the accuracy of the clash detection process. Some of the methods proposed include the use of clash detection software for automated detection in conjunction with the federated BIM model. Helm et al.~\cite{Helm2010} classified the algorithms used for building clash detection software into 4 categories: collision detection, ray-angle intersection, industry foundation classes (IFC), and open-source BIM servers. However, the authors stated that individually, each category lacks accuracy or speed. Thus, an IFC class detection that combines all four categories was recommended to balance the tradeoffs between accuracy and time constraints. Akponeware et al.~\cite{Akponeware2017} performed a detailed analysis of clash detection strategies, which revealed that a shared workspace that facilitates the co-creation of designs by stakeholders is the most effective method for tackling clash detection. This has been realized in common software such as REVIT (BIM 360 and Collaboration for Revit). They concluded by proposing an open work in progress (OWIP) section of the common design environment (CDE), which promotes simultaneous interdisciplinary works. However, their framework doesn't address privacy challenges among the stakeholders.

Beyond the detection of clashes, there is a significant challenge that exists in the industry as regards the identification and subsequent classification of these collisions \cite{RachaChahrour2021}. Several works of literature have attempted to classify clashes based on the severity of impact (cost over-run, time delay, rework, productivity loss, etc.) \cite{RachaChahrour2021,Hartmann2010,Akponeware2017}. Generally, the most commonly adopted classification is hard and soft clashes \cite{RachaChahrour2021,Akponeware2017}. Hard clashes exist where there is an overlap between members while soft clashes depend on clearance requirements \cite{RachaChahrour2021,Hu2019}. Clash detection software and tools identify a large number of clashes. However, some of the clashes are repeated, irrelevant globally or in terms of priority, or repeated false positives. Therefore, there is a need to filter the clashes detected based on the aforementioned classification and relevance to avoid waste of resources \cite{RachaChahrour2021,Hu2019,Hu2020}. In a bid to improve the quality of clash detection, Hu et al \cite{Hu2019} compared six different supervised machine learning algorithms which include: The J-48-based decision tree, random forest, Jrip-based rule method, binary logistic regression, naive Bayes and Bayesian network for clash filtering. The Jrip method was found to have performed the best out of all six methods. From this result, the authors proposed a clash filtering framework. 

In clash management, after the detection, classification, and filtering process, the stakeholders coordinate and communicate to resolve the identified clashes. This is the clash correction process. The level of automation of the clash correction process is lower compared with the clash detection process. Hu et al.~\cite{Hu2020} state that most research handles clashes independently ignoring the inter-dependencies and possibility of ripple effect that may occur from its resolution. Thus, the authors propose an optimized clash detection sequence, based on the clash network, which factors in the inter-dependencies. In conclusion, for automated clash detection, a minimum requirement of a level-2 BIM (earlier introduced in \cref{subsec: 4.3} is required \cite{Akponeware2017} and a holistic DT would address the fundamental challenge of stakeholder design silos \cite{Boje2020}.

\begin{figure}[h]
    \centering
    \includegraphics[width=0.4\textwidth]{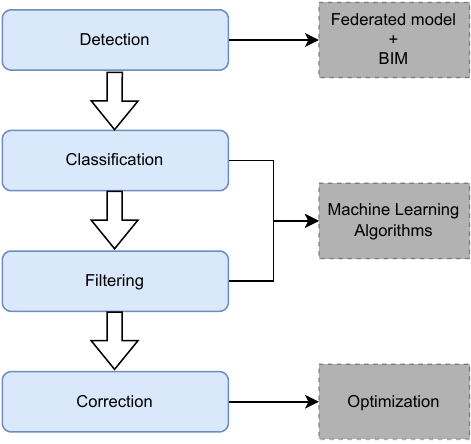}
    \caption{Clash management process and enabling methodology}
    \label{fig:Clash management process and techniques}
\end{figure}

\subsubsection{Optimized Construction Logistics and Scheduling}
A smooth interconnection between material supply, site operations, and construction personnel is vital for effective decision-making during the construction process. Efficient technologically infused logistics management has been scarcely adopted in the civil engineering industry \cite{Said2014,Boje2020,Greif2020}. This is mainly due to fragmented individual driving components and the difficulties encountered in the management and integration of large volumes of logistics data, which are usually dynamic during construction \cite{Said2014,Dave2016,Opoku2021}. Examples of these construction logistics data include site space availability, time and quantities of supply deliveries, and construction activities demand.

Over the past decade, optimization techniques have been one of the most employed methods to tackle the problem of construction management \cite{Dede2019,Greif2020, Said2014,Zhang2010,Afshar2009,Zhang2012,Aminbakhsh2016,ChoongwanKoo2015}. Said and El-Rayes \cite{Said2014} presented an automated multi-objective construction logistics optimization system (AMCLOS). AMCLOS is a framework that leverages BIM and optimization models for the seamless integration of construction data for optimal site decision-making. Dave et al.~\cite{Dave2016} focused on the improvement of the construction communication framework management. The authors emphasized the construction workflow as a key for optimizing construction processes, highlighting current construction management systems such as VisiLean and KanBIM. They concluded that the sufficient solution to the aforementioned problem is the enhancement of current management systems with IoT. Li et al.~\cite{Li2009} suggested virtual prototyping for the optimization of construction planning schedules through improved foresight via computer simulations. However, none of these solutions have been able to sufficiently address the integration and interoperability challenges of construction data. 

DT has been recognized as a potential solution capable of addressing the construction management challenges \cite{Greif2020,Sacks2020,Boje2020}. Although this potential has been recognized, there is limited literature to this effect, with most of the DT applications in the construction phase being tailored to structural systems integrity \cite{Opoku2021}. Notably, Greif et al.~\cite{Greif2020} proposed a decision support system (DSS) for silos-bulk materials storage space dispatch and replenishment leveraging DT capabilities and optimization techniques. The authors provide a lightweight digital silo twin which tackles construction logistics challenges specific to silos. These challenges include site work interruptions due to depleted fill levels, excessive and uncoordinated silos movement, and compromised operational planning due to poor quality of forecasts. Their DSS requires silos infused with sensors for data collection and fill monitoring. The data is subsequently optimized via a cluster-first-route-second heuristic optimization approach, to devise more efficient silos repositioning, routing, and replenishment. Sacks et al.~\cite{Sacks2020} presented a closed-loop construction workflow that utilizes DT information systems for construction control. Their proposed digital twin construction (DTC) comprises four core components and caters to both the construction process and product. The first component is the lean project production systems (LPS), which are employed for overall project monitoring and feedback loop. The second component is BIM: introduced as a core aspect of DT (see \Cref{sec: 4}) but was utilized majorly for information storage. Automated construction monitoring systems (sensors) are the third component of the proposed DT that provides input data. The final component is artificial intelligence. The details about its exact role in the DTC were vague.

The two applications by \cite{Greif2020}  and \cite{Sacks2020} are conceptual paradigms with little---just one existing implementation of the former---or no practical implementations. Furthermore, there was no indication in the two papers about the transition of their DTC framework into a DT for the operation and maintenance phase, corroborating the perceptive scarcity and the infancy of DT adoption in this sphere of construction.

\subsection{Remarks on DT Development and Application in the Construction Phase}
Taking cues from the adopted techniques for DT in construction as summarized, we realize the important roles of sensing and visualization technologies. In particular, laser scanning and photogrammetry are significantly used as a part of sensing methods for data capturing, and one of the many uses of BIM was visualization. At this point, we identify that for civil engineering projects, BIM remains central to the implementation of DT. Unlike the design phase, BIM is utilized in conjunction with other technological drivers for real-time monitoring, prediction, optimization, and control---all of which constitute the features of DT---in the construction phase. Notable uses of BIM for this phase include a comparative platform for automated progress monitoring via the Scan-vs-BIM techniques, visualization technology for construction workforce safety, federated model development and collaborative workspace provision for automated clash detection, and information storage for optimized construction logistics and scheduling.

We already noted IoT to be an important driver of real-time monitoring and thus its central role in DT development. However, a reflective observation from our reviews shows a segmented realization of IoT benefits in the AEC industry -- as most use cases are focused on two aspects of the entire project life-cycle in the AEC industry. First is sustainable design (energy management) which happens at the pre-construction stage through the rising success of BIM with cloud-based IoT services for simulation purposes as reported in \Cref{subsec: 4.1.4}. The second is operation and maintenance such as structural health monitoring which typically occurs post-construction stage. It is not far-fetched to realize that real-time data capturing through sensors and its immediate integration with virtual models (such as BIM) via the internet in a highly dynamic environment like construction sites poses a complex challenge. Hence the narrow applications of IoTs in the construction industry. Nonetheless, the advent of semantics, web ontologies, and linked data are technological pushers of IoT utilization in dynamic construction sites. 
Another important remark from the reported techniques is the rising use of computer vision as part of DT development for construction processes. In the next section, the notion of computer vision will be broadly introduced as it serves as one of the two main methods for structural health monitoring.

We conclude this section by stating that the automation of the various construction processes currently lacks control and feedback. We notice that manual inputs were significantly required towards the end of any automated process to drive the next input actions. For example, we reported above that the level of automation in clash correction remains lowest when compared to the earlier three steps of clash detection, classification, and filtering. Also, one of the highlighted limitations in construction safety is manual safety measures update. These expositions show the status quo of DT implementation and the need for increased research focus on driving fully automated processes including control and feedback in the construction phase.

\section{Operations and Maintenance: SHM and DT}
\label{sec: 6}
 We commence this section by delineating the subtle differences between DT and SHM. Gharehbaghi et al.~\cite{Gharehbaghi2022} segmented SHM into two broad categories termed diagnosis and prognosis. From several kinds of SHM research conducted over the years, some of which are later reported, one can infer that major work has been done on the diagnostic parts \cite{Rytter1993} (damage detection, localization, and evaluation) with comparatively low progress on the prognosis aspect of SHM such as residual life prediction. Nonetheless, the advent of several miniaturized or sensor-based SHM systems directly attached to modern structures has paved the way for real-time monitoring of the structures via model updating and parameter estimations, otherwise known as system identification \cite{Zhu2020}. With these tools in place, the prognosis aspect of SHM and more features become attainable including prediction, learning, management, optimization, and other context-specific functionalities---all of which are encompassed in a virtual duplicate of a physical infrastructure tagged DT \cite{Wagg2020}. Thus, DT can be viewed as a global system that can perform complete SHM in real time.

As similarly done for the other phases, we attempt to avail readers of the fundamental concepts of SHM in the O\&M phase. It should be noted that myriads of reviews on SHMs have been done. As such, we quickly focus on relevant techniques that come together towards DT development for SHM. Notably, we establish the evolving methodologies of SHM including physics-based SHM methods, purely data-driven approaches using either raw data or extracted features from engineered data, or a combination of both, alongside techniques such as model update and regularizations to tackle the inverse problem formulation of SHM. On this premise, we introduce the requirements of SHM-based DT and existing mathematical formulations of the same toward the attainment of a DT for SHM.

\subsection{SHM: Overview and Enabling Technologies} 
\label{subsec: 6.1}
SHM can be technically defined as the process of deploying methods for damage detection and analysis via the evaluation of the current mechanical state of the infrastructure, all geared towards condition-based awareness of infrastructure and its continued optimal performance. Traditionally, SHM was implemented on a scheduled basis via expert visual inspections. This method is very inefficient due to human limitations and the lack of visibility of some damages, resulting in the wastage of resources. Therefore, recent technological advancements have resulted in an automated SHM process. Generally, SHM can be implemented globally or locally. Global monitoring involves monitoring damage in the entire system, which is usually employed for older structures and when certain structural parts are inaccessible. Local monitoring involves the evaluation of specific parameters in the structure, which is usually employed for non-complex and modern constructions \cite{Gharehbaghi2022,Alonso2018,AlHamaydeh2022,Liu2021,Chen2018}.  

SHM processes can be described in three general steps, i.e., data acquisition, data processing, and data interpretation. Data acquisition involves observing relevant parameters, often using sensors (see \Cref{sec: 5.1}). These sensors may be contact, non-contact, wireless, and give active (actuators) or passive---more commonly employed---feedback. The data processing phase may be signal processing techniques such as the Fourier transform, short-time Fourier transform (STFT), wavelet transforms (WTs), empirical mode decomposition (EMD), and Hilbert-Huang transform (HHT). It also involves the normalization of data, filtering, cleansing, transmission, management, and storage \cite{Farrar2007,Gharehbaghi2022,AlHamaydeh2022,Goyal2016}. Finally, data interpretation involves the damage assessment (diagnosis and prognosis). Some commonly employed techniques include support vector machines, fuzzy logic, deep neural networks, and Bayesian classifiers \cite{Alonso2018}.

The taxonomy of SHM can be also described in terms of the techniques employed. The most commonly employed SHM technique is the vibration-based method, which is dependent on the parameter to be observed in the structure. Structures may be subjected to forced vibration, caused by applied external forces or ambient excitation. Ambient excitation is preferred as they do not require external shakers which may damage the structure further. These excitations result in vibrations which allow certain parameters to be monitored. These parameters can be either static or dynamic. The dynamic parameters include the natural frequencies, modal shapes, mode curvature, damping, flexibility, and stiffness matrices, which are more sensitive to changes than static parameters such as strain or stress \cite{Gharehbaghi2022,AlHamaydeh2022}. SHM techniques may also be image-based, particularly the use of computer vision \cite{Alonso2018} (see \Cref{subsec:6.2}). Another classification is model-based or signal-based SHM. Model-based SHM involves the observation of model parameters or errors for damage detection. These models may be physics-based or data-driven (see \cref{subsec:6.3}). For signal-based SHM, changes in system characteristics such as means, variances, and spectral information can be evaluated by using signal analysis for feature extraction from measurement data. Damage detection decisions are thus made based on the data analysis via time-domain, frequency-domain, spatial-domain, and time-frequency domain methods. This is also called feature-based SHM \cite{Goyal2016,Gharehbaghi2022,Gallet2022,Farrar2007}. A closely related SHM perspective is that of pattern recognition. Although discussed independently due to its popular adoption, it is a data-driven SHM method. These methods rely on algorithms for the classification and assessment of damage. The algorithms are dominantly classified into three: supervised learning for data that contains prior knowledge about various damage scenarios, unsupervised learning for data that does not contain examples from the structure; and the last group is the reinforcement learning algorithms \cite{Flah2021}.

Mathematically, SHM is generally posed as an inverse problem \cite{Gallet2022,Gharehbaghi2022}. Simply put, the inverse problem involves finding the cause of a problem from the effect of the problem. This involves using measurements from the structures as inputs of a model to generate the characteristics of the structure as an output. Inverse problems are generally ill-posed and sensitive to perturbations. A forward problem can be written as \cite{Gallet2022}:
\begin{align}
    d = U(\theta) + \delta{d}
\end{align}
where $\theta$ is the structural parameter, $U$ is the forward model of the structure, $d$ is the measured data and $\delta d$ is the error term, representing errors stemming from different sources such as measurement inaccuracies or model simulation inaccuracies. In this case, $\theta$, which is sensitive to perturbations, is to be evaluated from $d$. However, the instability and ill-posedness of inverse problems prevent its formulation by simply inverting the forward problem. Several classes of methods exist for inverse problem formulation that address the above challenges, one of which is the direct computation of $\theta$ from $d$:
 \begin{equation}
    U^\dag (d) \approx \theta
\end{equation}
where $U^\dag$ is an inverse mapping. This class is quite challenging to execute largely due to its problem dependency. A generally applicable method is by constructing the inverse problem as an optimization problem given by:
 \begin{equation} 
    \theta^* = \underset{\theta}{\argmin} \, \frac{1}{2}||U(\theta - d)||_2^2 + \alpha{R(\theta)}.
\end{equation}
The first part of the equation enforces the reconstructed problem fit the measured data, while the second part ${R(\theta)}$ is the regularization term. The regularization term provides stability to the constructed equation by enforcing a unique solution for each set of measurements while also preventing the solution from overfitting the measurement noise. The parameter $\alpha>0$ balances out the two parts.
Gallet et al.~\cite{Gallet2022} provided different views on SHM as an inverse problem. Notably, they divided SHM techniques as deterministic or probabilistic. Another is the static or dynamic SHM inverse problem, where model updating (see \Cref{subsec: 6.4}) was advanced as a good approach for handling the latter. Other approaches include guided wave-based modalities \cite{Ochieng2018}, inverse filtering, and Bayesian approaches. Computer vision inverse problems in SHM were also presented (see \Cref{subsec: 6.2}). The subsequent subsections further expatiate the prevalent SHM techniques in civil engineering, their advantages and challenges, and their relationship with DT.

\begin{figure}[t]
    \centering
    \includegraphics[width=0.45\textwidth]{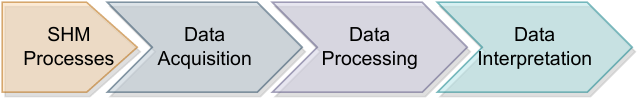}
    \caption{SHM methodology}
    \label{fig:SHM Methodology}
\end{figure}

\begin{figure}[h]
    \centering
    \includegraphics[width=0.2\textwidth]{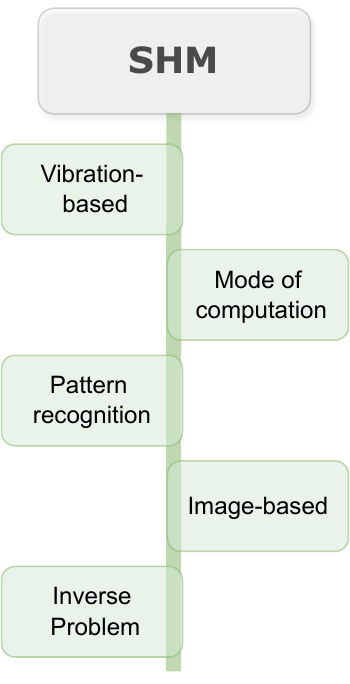}
    \caption{SHM classification perspectives}
    \label{fig:SHM Classification Perspectives}
\end{figure}

\subsection{Computer Vision: Image Processing and Deep Learning Based Approaches}
\label{subsec:6.2}
The use of contact-based sensors for SHM presents several challenges such as technical installation requirements, limited access due to occlusions, electromagnetic interference, and global view restrictions for system identification. As such, computer vision, which offers non-contact, long-distance, improved precision, and reduced occlusions, is being employed as a faster solution for SHM \cite{Jahanshahi2011,Lee2012,Koch2014,Ye2016}. In this subsection, we present the fundamental concepts of computer vision and their translations for data-driven SHM methods. We decided to provide these expositions due to their rising adoptions as online models in the overall DT system for SHM, as presented in \Cref{subsec: 6.4}.

\subsubsection{Computer Vision for SHM: Motivations and Conceptual Overview}
\label{subsec: 6.2}
Computer vision includes the combined use of a vision system (imaging acquisition devices from cameras to laser scanners), a computer, and image processing software which is integrated with specific computational algorithms to automatically extract useful information from image datasets \cite{Gharehbaghi2022,Ye2016}. Gharehbaghi et al.~\cite{Gharehbaghi2022} highlighted two major groups of computer vision, especially in the context of SHM, which include Machine Learning (ML) or image processing techniques and Deep Learning (DL) approaches. While ML and DL are similar in their ability to classify image datasets, their nuances make them preferable in different contexts. ML techniques involve statistical methods, and feature selection and engineering with typically small model sizes that are applied to structured/tabular data. DL techniques involve deeply connected neural networks with large (overparameterized) model sizes that are applied to semi- and unstructured data (images, texts). Noting that training DL models from scratch is expensive, transfer learning which typically reuses a piece of an already trained model for related tasks has become an important technique for diverse applications.

\begin{figure}[h]
    \centering
    \includegraphics[width=0.5\textwidth]{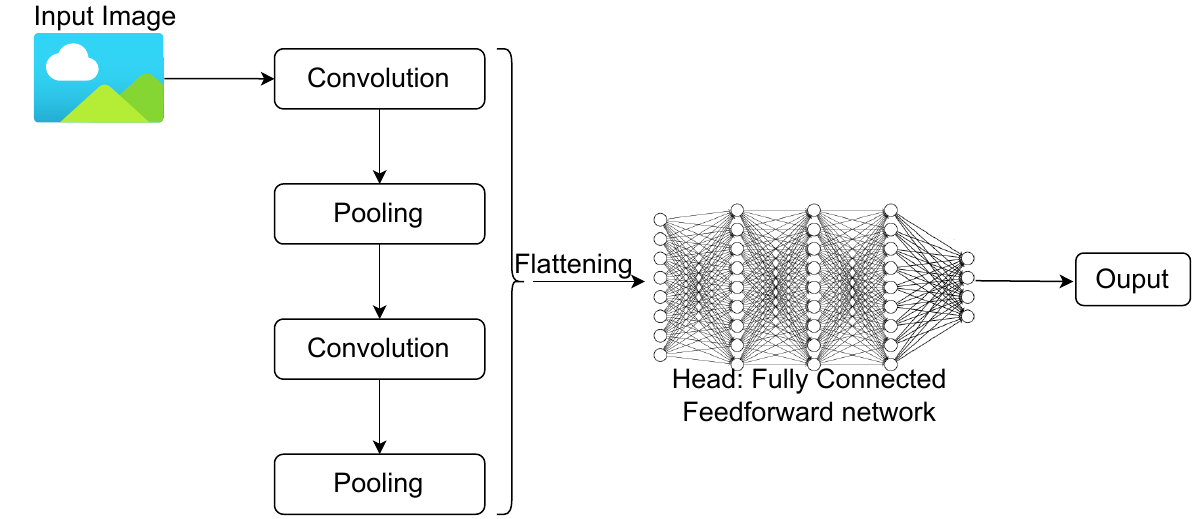}
    \caption{A simplified representation  of full convolutional neural networks}
    \label{fig:Full_CNN}
\end{figure}

Although ML techniques such as edge detection filters and morphological features have been considerably good in image classification tasks, the dimensions of fully connected neural networks (FCNN) needed for the increasingly large number of input variables from image datasets pose a great challenge. However, the non-unique nature of images (i.e., approximate representations of natural objects by different pixel value combinations) offers a robust advantage for big data processing, inspiring the development of convolutional neural networks (CNN). CNN compresses a dense FCNN by sliding a reduced linear kernel over a fully connected node, followed by pooling (progressively scaled patterns at different resolution levels), and/or flattening for feature processing (Fig.~\ref{fig:Full_CNN}). Capitalizing on the pixel-level representations of images, variants of CNNs such as deep CNNs (DCNNs)) have been developed to go beyond classification tasks. Notable mentions include object detection via you only look once (YOLO) or single shot detector (SSD) networks, and segmentation via transposed convolution such as deconvolutional networks (DeconvNet) and semantic segmentation model (SegNet). The idea behind these advancements can be simplified to a combination of classification with localization to extract multiple instances from the same class of objects \cite{Girshick2014,Redmon2016,DeBrabandere2017,Perazzi2017}.  

Spencer Jr et al. \cite{SpencerJr2019} summarized the recent advanced techniques for computer vision into supervised end-to-end learning including CNN and DCNN briefly described above, unsupervised learning comprising probabilistic structures and CNN unsupervised learning, and optical flow techniques. Optical flow techniques are predominantly employed in the dynamical monitoring aspect of infrastructural condition assessment, which involves quantitative measurement to infer the current states of structural systems \cite{Horn1981}.

\subsubsection{Computer Vision for SHM: Applications}
The aforementioned advances in computer vision have been applied to the diagnostic aspect of SHM, with limited extension to the prognosis aspect of SHM. For ML-based crack detection, which is prominently employed, Salman et al.~\cite{Salman2013} utilized Gabor filtering to automatically distinguish cracks in images with a detection precision of 95\%. SpencerJr et al.~\cite{SpencerJr2019} categorized the use of hand-crafted filters and ML classifiers as ``heuristic feature-extraction methods'' and particularly reviewed four different methods for concrete crack detection, which comprise edge detection filtering, multiple sequential image filtering, median filtering, and depth and 3D information. A robust review on crack detection using image processing techniques can be found in \cite{Mohan2018}. We note that feature extractions in ML techniques are integral for SHM, as we cannot apply raw data---which are often misleading---for damage assessment \cite{Long2014}. However, feature extractions such as hand-crafted features are usually expensive, sub-optimal, and inextensible to other structures \cite{Yuan2020}. These difficulties suggest the use of data in their raw form for automated damage detection, which is inapplicable via ML. In this regard, DL methods have been widely adopted due to their ability to use data in their raw form and learn features automatically. 

DL techniques have also been well-adopted for SHM due to the proliferation of cameras and significantly improved computation capabilities such as graphical processing units (GPU), and associated complexities with big data. Zhang et al.~\cite{Zhang2017} developed CrackNet, a CNN trained on 1,800 images using two GPU devices for asphalt crack detection. Worden and Dulieu-Barton \cite{Worden2004} combined DL with Bayesian optimization \cite{Do2023,Do2024} to inspect the conditions of a reinforced concrete bridge subjected to an earthquake. Hoskere et al.~\cite{Hoskere2018} utilized FCNN to detect damage and identify building components to establish damage presence, location, type, and scene building (SB) information, all geared toward structural component recognition and damage detection. On another end, optical flow algorithms are used for dynamic monitoring while digital image correlation (DIC) has been reportedly successful for static applications.

\subsection{Simulation-Based SHM}
\label{subsec:6.3}

The inference step of SHM involves the development of a reference for systems monitoring, which is typically achieved through a physics-based and/or data-driven approach. Taddei et al.~\cite{Taddei2018} noted that the difference in these approaches can be established through adopted mechanisms at the online stage of the monitoring process. The authors argued that the offline stage of a data-driven method might include physics-based synthetic experiments for data acquisition and thus limit the distinctiveness at that stage. We, however, state that the body of literature has carefully distinguished the use of data-driven approaches based on physics-based synthetic experimental data to be a physics-informed data-driven approach, otherwise tagged as hybrid-based simulation techniques. As such, we present three main computational approaches for SHM, including their conceptual overview, applications, and limitations, giving place for DT as a more efficient tool for SHM.

\subsubsection{Physics-based Simulation Techniques}
Physics-based simulation techniques are based on physical laws governing the behavior of the structures of interest, and aiding the extraction of features from measured sensor data for damage detection and evolution. The physical models are presented by differential equations, which are built and calibrated during the offline stage of monitoring processes \cite{Yuan2020,Taddei2018}. During the online stage, the model is updated based on newly measured data and instantaneously recomputed to get the current state of the structure.

In a broad sense, analytical and numerical techniques have been developed to solve differential equations extended to SHM. While frequencies and mode shapes are commonly employed as modal parameters in SHM methods, the former cannot be used to detect local damages since they are global indicators (i.e., scalar), hence the latter which are local indicators (i.e, vectors with local information) are often used. The extracted mode shapes are then compared for both damaged and undamaged stages using damage indices such as modal assurance criterion (MAC) or coordinate modal assurance criterion (COMAC) for damage localization and evaluation. Another notable numerical approach for local damage identification is the use of finite element (FE) model updating to construct baseline structural models more accurately before comparisons as a basis for reconstruction \cite{Yang2021,Friswell1995,Sanayei2015,Suzuki2017}. Precisely, the stiffness matrix of the FE model is updated so that the frequencies and mode shapes match the measured ones from the monitoring system, which are then used to infer the damage location and evaluate its severity (i.e., stiffness loss). The FE model updating is generalized as a constrained optimization problem given by: 
\begin{align}
    \min \left\| \sum_{i} w_i(\lambda_{\mathrm{FE},i}(x_k) - \lambda_{i} )\right\|^{2}_{2} \quad \text{such that} \quad x_{lk} \leq x_{k} \leq x_{uk}
\end{align}
where $\lambda_{\mathrm{FE},i}(x_k)$ is the $i$th frequency or mode shape obtained by FE model using design parameters $x_k$, $\lambda_i$ is the measured $i$th frequency or mode shape, $w_i$ is the weight factor in the range of 0 to 1, $x_{lk}$ and $x_{uk}$ are the upper and lower bounds on the $k$th design variable, respectively. Since the number of measurements is usually less than the number of unknown parameters, FE model updating is ill-conditioned. Additional term and regularization parameters are adopted to circumvent such bottlenecks \cite{Yang2021}. The Tikhonov regularization is a commonly adopted technique in practical scenarios. \Cref{subsec: 6.4} will provide more details on model updating as it relates to the creation of DT for SHM. Other physics-based SHM techniques as summarized by \cite{Yuan2020} include but not limited to modal strain energy \cite{Kim2003,Shi2000}, modal flexibility \cite{Pandey1994,Jaishi2006}, wavelet transform \cite{Staszewski1997,Ovanesova2004}, damage locating vector \cite{Bernal2002}, Hilbert-Huang transform \cite{Yang2004}.

Despite the success of numerical methods for increasingly complex problems, physics-based techniques are inflexible to dynamically updated models, as changes in the source functions and/or initial/boundary conditions require a total re-computation of the model solution, birthing the curse of dimensionality \cite{Yuan2020}. Zhu et al.~\cite{Zhu2020} also reported identification uncertainties arising from uncontrollable and unknown operational input and environmental variability as associated challenges of the physics-based SHM approaches.  As such, other techniques have been developed from a purely data-driven perspective or a physics-informed (hybrid) approach. Also to tackle the effects of measurement noise and model errors, statistical methods based on Bayesian inference have been developed, which are particularly useful for DT creation for SHM as reported in \Cref{subsec: 6.4}.

\subsubsection{Data-driven Simulation Techniques}
\label{subsecsec:6.3.2}
Data-driven approaches are primarily based on statistical pattern recognition from the dataset, eliminating the need for FE model updating and re-computation during the online phase of SHM. This pattern recognition is however done by a trained classifier, necessitating the use of ML and DL frameworks. The ML and DL techniques introduced in \Cref{subsec: 6.2} are mostly applied here but are not limited to image or video data. Concisely, the data-driven approach comprises an offline expensive phase including data acquisition, feature extraction, and training the classifier, and an online phase that utilizes the trained classifier for SHM \cite{Bigoni2020}. For SHM, the pattern recognition strategy is often regarded as ``anomaly detection'' with a fundamental framework of checking the conformity of newly generated datasets to the features extracted from the baseline system. The use of ML for anomaly detection is not a trivial task since they are based on binarization. In other words, the use of supervised ML techniques for damage detection will require data from both the undamaged baseline and the anticipated damaged state of the structure. This is limited for real-life applications as it is impractical to generate datasets that reflect all possible damage scenarios of a structure \cite{Taddei2018,Long2014}. 

To circumvent the aforementioned difficulties, One-class classification (outlier detection or novelty detection) is often employed in SHM, where labeled data from the undamaged state are used in the training phase and unlabeled data from both the undamaged and damaged states are used in the test phase to detect anomalies employing an online novelty score \cite{Pimentel2014,Goldstein2016}. Three well-known one-class classification strategies include Isolation Forest \cite{Liu2008}, Local Outlier Detector \cite{Breunig2000}, and the One-Class Support Vector Machine (OCSVM), which is commonly adopted for SHM \cite{Long2014,Das2007, Anaissi2017,Bigoni2020}. The OCSVM is derived from the binary classification: Support Vector Machine (SVM). Unlike the binary SVM, the OCSVM does not have second-class data but instead, the decision boundary (hyperplane) separates the data from the origin, as a proxy for the second class, which is formulated as a convex optimization problem.
Long et al.~\cite{Long2014} and Schoelkopf et al.~\cite{Schoelkopf2001} provided reviews on OCSVM.
Das et al.~\cite{Das2007} used OCSVM as a signal-processing technique in a wave-based approach for characterizing defect states in composite laminates in an anisotropic medium. On a different spectrum, compression techniques were advanced due to the established need for memory-efficient SHM techniques that can be directly used on the sensors or the gateways \cite{Thaprasop2021,Siddiqui2020,Zonzini2020,Moallemi2021}. Notably, Principal Component Analysis (PCA) and Autoencoder are used to transform high-dimensional correlated data into low-dimensional correlated data \cite{Cui2019}. Moallemi et al.~\cite{Moallemi2021} utilized autoencoders as a data-driven approach to detect anomalies in a bridge structure.

Data-driven approaches for SHM are greatly limited by the prohibitive cost of offline data acquisition. We once again reemphasize that it is impractical to have access to datasets that are representative of all the possible damage scenarios of a structure for supervised ML classification. Although we earlier presented one-class unsupervised ML methods such as OCVSM for SHM, such methods are only successful for the lower levels of damage diagnosis (only detecting damage presence \cite{Yuan2020,Rytter1993, Taddei2018}) let alone establishing damage prognosis (such as damage severity or life-cycle estimation). We recall that the limitations of physics-based approaches---particularly as inverse problems---occur mostly at the online stage, while that of data-driven, as presented herein, occurs mostly at the offline stage. It is thus plausible to hybridize these two approaches as a means for improved SHM, termed hybrid-based simulation techniques.

\subsubsection{Hybrid Simulation Techniques}
\label{subsubsec: 6.3.3}
The concept of regularization for solving inverse problems is also viewed in the data-driven sense as a guided training process using domain knowledge \cite{Yuan2020}. Yuan et al.~\cite{Yuan2020} report that the integration of domain knowledge for model regularization can either be from a mathematical model governed by PDEs (tagged physics-informed machine techniques) or from expert human knowledge through techniques like transfer learning and network architecture customization. Among the several ways \cite{Ritto2021} of creating physics-informed machine learning, a commonly adopted one for SHM is the synthetic creation of data from physics-based models for training ML classifiers. In addition to sufficient data creation, the use of a physics-based model aids the interpretability of ML tools and permits further analysis beyond damage detection such as establishing damage location and severity.

The use of physics-based simulations in the data-driven approach is however limited by the computational burden in solving parameterized PDEs for many values of parameters. For improved efficiency, the concepts of surrogate modeling \cite{Razavi2012, Chua2021} and adaptive sampling \cite{Basudhar2008,Basudhar2008a} were introduced. However, both approaches are still limited by the curse of dimensionality and are mostly applied to static data. Another limitation is model errors from the approximate representations given by those techniques \cite{Taddei2018,Yadav2024}. Parametric model order reduction (pMOR) and definition of bounds (uncertainty quantification) have been adopted to address the curse of dimensionality and model errors, respectively. A commonly adopted pMOR technique is the Reduced-Basis (RB) method \cite{Hesthaven2016,Quarteroni2015}: a projection-based method with the central idea of using a linear combination of suitable basis functions derived from a high-fidelity problem to reconstruct a solution for a new parameter \cite{Bigoni2020}. The RB method is based on the reformulation of several problems in the following form: for a given value of $\mu$, find $u(\mu) \in X$ such that
\begin{equation}
    F(u(\mu); \mu) = 0
\end{equation}
where $X$ is a function space, $\mu \in \mathbb{R}^{p}$ is a vector of parameters representing materials properties, time, geometric properties, etc., and $F$ is a mapping defined over $X \times \mathbb{R}^{p}$. Provided some initial values $u(\mu_i)$ for the function, $u(\mu)$ can be nicely predicted for a generic value $\mu$ \cite{Maday2002}. For example, we can predict displacements based on some parametric evolution such as geometry or time (typically called snapshots), where we identify the dominant basis functions for approximation. Through this, repetitive use of the underlying model becomes feasible enabling applications for real-time control and optimization and in turn diagnostic and prognostic SHM \cite{Maday2002}. However, the RB method consists of a potentially expensive offline stage to generate the reduced basis $\Phi_r$ (a low-rank truncation of the data) and an online stage typically based on Galerkin projection given by:
\begin{equation}
    u(\mu) = \Phi_{r} a(\mu)
\end{equation}
where $a(\mu)$ are to be determined from the dynamics on the $\Phi_r$ subspace.
Another approach to pMOR is to use a reduced subspace $\Phi_r(\mu)$ that adapts with the parameters,
typically via interpolation~\cite{Zhang2022}.
Compared with using a global basis, this approach can achieve the same accuracy
with a much smaller reduced model, which is very attractive for online DT applications.

Bigoni et al.~\cite{Bigoni2020} employed a data-driven approach through physics-based data generations (making their approach hybrid) obtained from solving parametric PDEs using pMOR based on Galerkin projection, and the reconstruction of time signals using the Weeks method (a numerical inverse Laplace transform). OCSVM was then used as the classifier using newly synthesized data obtained from the solution of the parametric PDEs for unseen input parameters. Their proposed approach was successful for damage detection and localization with reported extension to 2D and 3D DT frameworks.

Uncertainty quantification has been implied in the reported applications of SHM. At this point, we briefly introduce Bayesian methods which have been commonly adopted for mitigating model errors' impact on damage detection and localization. Bayesian methods are probabilistic methods that use prior information from experiments or experience to give the posterior probability of uncertainties in damage identification and localization. Bayesian methods intrinsically regularize models using the associated probability distributions. Their successful application is however hinged on the accuracy of the prior information \cite{Yang2021}. To generate accurate prior with limited data availability, sparse Bayesian learning \cite{Tipping2001,Wipf2004,Zhang2011a} has been developed for constructing parameterized prior based on sparse data \cite{Yang2021}. For a given structure response $R$, the posterior probability of parameter $\theta$ is given by Bayes rule:
\begin{equation}
    p(\theta | R) = \frac{p(R|\theta)p(\theta)}{p(R)}
\end{equation}
where $p(R|\theta)$ is the probability of $R$ given $\theta$, $p(\theta)$ is the prior probability of $\theta$, $p(R)$ is a normalizing constant, and $p(\theta|R)$ is the posterior probability distribution of $\theta$ given $R$. Bayesian methods are particularly useful for the development of DT for SHM and further details will be provided in \Cref{subsec: 6.4}.

Another promising approach is the use of ANN to solve differential equations. ANN are utilized to learn complex non-linear mappings to swiftly handle slight variations in models. With the advent of GPUs and TensorFlow (DL framework with embedded automatic differentiation), research in this area has considerably increased for solving nonlinear PDE (see \cite{Raissi2019}). Yuan et al.~\cite{Yuan2020} demonstrated the promise of this approach to solve both forward and inverse problems for beam vibration modeling and displacement reconstruction respectively. For the forward problem, authors sequentially employed the collocation method in segmented short-time windows across two training phases successively without and with the loss function. The ANN-solving mechanism was then extended to the inverse problem involving displacement field reconstruction from a sparse array of data.

This section has introduced the three-phase techniques commonly adopted for SHM. It is the authors' opinion that such expositions set the tone to introduce the conceptual advancements of SHM using DT both from implementation and applicability perspectives presented in the subsequent sections.

\subsection{SHM + DT: Model Update and Uncertainty Quantification}
\label{subsec: 6.4}

\begin{figure}[t]
    \centering
    \includegraphics[width=0.5\textwidth]{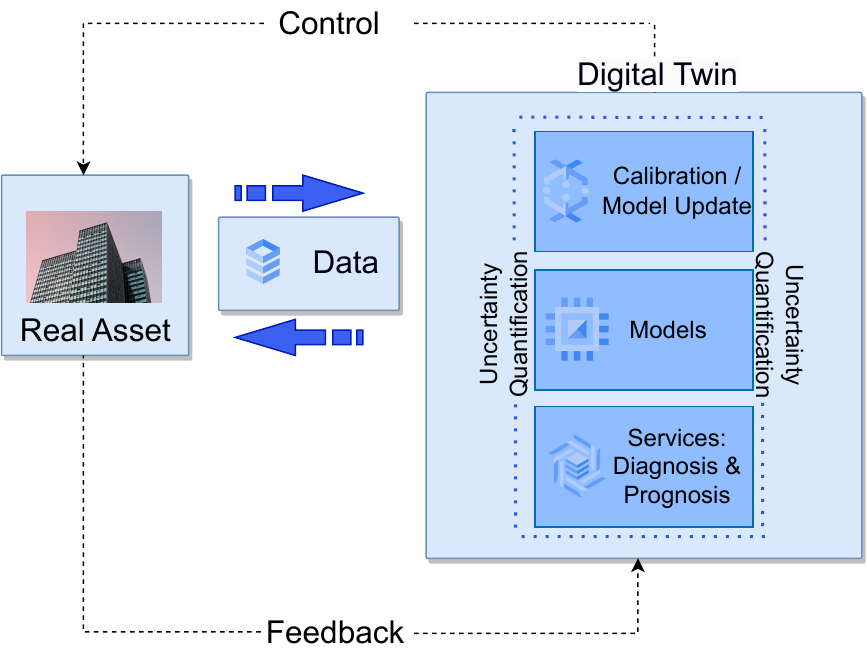}
    \caption{DT for structural health monitoring. \textnormal{\textit{Requisite components include a real asset, virtual representation with computational models calibrated for the physical asset, real-time bi-directional data exchange, feedback, and control for prognostic and diagnostic services. Uncertainty quantification is essential in the DT components.}}}
    \label{fig: SHM+DT}
\end{figure}

We emphasize that the creation of DT for SHM purposes is made possible by the rising success in model updating, data augmentation~\cite{Zhang2020}, and uncertainty quantification~\cite{Zhang2021}. In the proposed DT framework developed by Ritto et al.~\cite{Ritto2021}, a DT goes beyond a computational model of underlying physical structures but accounts for uncertainties, calibrated with data measured from the physical twin, evolving with time to reflect the current states of the asset and life-cycle estimation. In a bid to distinguish between a validated model and DT, Wagg et al.~\cite{Wagg2020} emphasized time-evolution and extended use of DT as characteristic features of DT, noting model updates and uncertainty quantification as key functionalities. In fact, the authors interchangeably used model updating and uncertainty quantification, further revealing their inherent connection. Although several techniques such as regression and Bayesian methods have been established for uncertainty quantification, their applications in DT are complicated by increasing complexities from multiphysics or multiscale modeling arising in several system-level predictions. Approaches such as data-augmented modeling or surrogate modeling were highlighted to circumvent the challenges of uncertainty quantification for DT. The authors still advocated the need for more efficient methods for estimating uncertainty propagation in DT.

In a later work by Zhu et al.~\cite{Zhu2020}, a model updating strategy was developed to alleviate the computational burdens associated with the earlier introduced strategies. They proposed the use of an intermediate model to eliminate the need for continuous model updating which might be unnecessary for unaltered systems. The three-level strategy comprises an intermediate model which can be any surrogate with identifiable parameters from SHM data, divergence analysis for conditioning model update based on the quantified level of system changes, and a model update using Bayesian methods to propagate the uncertainty between the measured data, intermediate model, and the system model. The authors' novel contribution is embedded in the divergence analysis step which is also in two parts: the first is to check if measured changes are due to environmental conditions or system changes, and the second step proceeds if the first check is attributed to system changes such that this change is compared to an established threshold to necessitate the model update. The authors demonstrated their proposed strategy for a vibration-based system identification method. For this, a modal model was employed as the intermediate model, where modal analysis was conducted to get the posterior distribution of the parameter $\omega$ of the modal model given the measured data $D_i$, denoted by $p(\omega|D_i)$. A divergence analysis is then conducted such that a discrepancy $d_i$ between the intermediate model and reference model is quantified and compared with the tolerance discrepancy $d_\mathrm{tol}$ as a determinant for model update or not. The model update itself utilized a Bayesian framework to account for the identification uncertainty of the model parameters given by:

\begin{equation}
   p(\theta|D_i) = p(\theta) \int \frac{p(\omega|\theta)}{p(\omega)} p(\omega|D_i)d\omega
\end{equation}
where $\theta$ is the system parameters which are updated by the posterior distribution given above. At this point, we follow suit with other researchers and emphasize that uncertainty quantification remains an important link for the creation of DT and it must be accounted for from the calibration to the prediction and optimization of a DT as shown in \Cref{fig: SHM+DT}. The next section further highlights the stronghold of uncertainty quantification for DT creation.

\subsection{Existing Mathematical Abstractions for DT}
\label{subsec: 6.5}
\subsubsection{Probabilistic Graphical Model}
A well-received mathematical abstraction of DT proposed by Kapteyn et al.~\cite{Kapteyn2021} involves six representative quantities: physical state ($s \in \mathbf{S}$), observational data ($o \in \mathbf{O}$), control inputs ($u \in \mathbf{U}$), digital state ($d \in \mathbf{D}$), quantities of interest ($q \in \mathbf{Q}$), and reward ($r \in \mathbf{R}$). A crucial selling point of their mathematical abstraction is its utilization for developing mathematical models for DT that are currently being applied to structural systems for quantification and validation. One of such is the probabilistic graphical model (PGM) for DT  which encodes the interaction and evolution of the six representative quantities of the asset-twin system. The encoding captures the data-decisions-flow (ranging from sensing, assimilation, inference, and action). Statistically, the model is built upon a dynamic Bayesian network with the inclusion of decision nodes, which are time-discrete random variables representing each quantity.

The probabilistic construct of the model can be linked to the nodal representation of discrete-time quantities as random variables; for example, the digital state $D_t$ at time $t$ is defined as a random variable distributed as $p(D_t)$. The graphical framework enables a principled approach for calibration, data assimilation, and feedback control loop of the DT leading to important features like prediction, planning, and uncertainty quantification. The conditional independence structure of the graph permits the factorization of joint distributions of the model variables. In particular, the factorization serves two purposes which include revealing the interactions to be minimally included in the DT computational models and providing a basis for efficiently derived Bayesian inference algorithms combined with modeled interactions for specific DT actions. An example factorization given by \cite{Kapteyn2021} is shown below:
\begin{align}
 p\left((D_t, Q_t, R_t)_{t=0}^{t_c} \big\vert (o_t, u_t)_{t=0}^{t_c}\right)
 = \prod_{t=0}^{t_c} \left[ \phi^{\mathrm{update}}_{t} \phi^{\mathrm{QoI}}_{t} \phi^{\mathrm{evaluation}}_{t} \right]
\end{align}
where the factors are defined as
\begin{align*}
    \phi^{\mathrm{update}}_{t} &= p(D_t|D_t-1, U_t-1 = u_t-1, O_t = o_t), \\
    \phi^{\mathrm{QoI}}_{t} &= p(Q_t|D_t), \\
    \phi^{\mathrm{evaluation}}_{t} &= p(R_t|D_t, U_t = u_t, O_t = o_t).
\end{align*}
In \Cref{subsec: 6.6}, we report an application of the PGM for SHM of civil engineering structures by  Torzonni et al.~\cite{Torzoni2024}.

\subsubsection{Digital Mirrors and Virtualization}
Before the development of the probabilistic graphic framework presented above, Worden et al.\cite{Worden2020} made an attempt to provide a mathematical framework for DT (or digital mirrors). To present their mathematical framework, the authors reserved the use of the term digital twins for digital mirrors in the concept of context-dependent virtualization. They argued that existing ``DTs'' are context-dependent and as such it is not always the case to fully model a physical system, and that most models reflect specific features or parts of a bigger scope in a system. In addition, the literary definition of the mirror as an object that faithfully reflects a mirror-facing object (i.e., context-dependent) flows with their argument. This paper will however not separate the rhetoric of mirror from twin but rather present their mathematical underpinnings for DT modeling and applications. The term ``digital twin'' will be adopted throughout the rest of this paper for the sake of generalizations.

Similar to the PGM's representative quantities, the authors attempted to define the physical object of the system $S$, environment $E$, context $C$, schedule $W_C$, and associated schedule's test $T^{C}_{W}$ to present the mathematical abstraction of digital twin (mirror). The physical system $S$ will be represented with a state vector $\mathbf{s}(t) = \{s_{1}(t), \cdots, s_{N_S}(t)\}$ across $N_S$ time-specific measurements to characterize its states, and the environment $E$ which is viewed as an influence on $S$ is represented by an environmental state vector $\mathbf{e}(t) = \{e_{1}(t), \cdots, e_{N_E}(t)\}$. The desired system aspect (context $C$) is a set of environmental state variables $C = \{e^{C}_{i} \in E, s^{C}_{j} \in \mathbf{s} : i,j\}$, where $e^{C}_{i}$ represents environmental context and $s^{C}_{j}$ represents response or predictive context. The schedule $W_C$ is a time series set given by $\{\mathbf{e}^{C}_{W}(t_i) : i = 1, \cdots, N_i, t_i \in [0, T]\}$, where $\{t_i\}$ could be continuous or discrete. The response $\mathbf{r}^{C}_{W}(t)$ to $W_C$ is thus represented as the measurement sequence from structural test results that are based on schedule inputs, which is represented by the functional 
\begin{align}
\mathbf{r}^{C}_{W}(t) = S[\mathbf{e}^{C}_{W}(t) \equiv W_C]. 
\end{align}
The associated test schedule can be defined as $T^{C}_{W} = \{\mathbf{e}^C_{W},\mathbf{r}^C_{W}\}$, with data captured for both training $D_\text{train}$ and testing $D_\text{test}$.

A model $M^C$ is defined to predict the behavior of the physical object $S$ for any schedule specific to the context $C$. As such, they deduced a simulation for a context $C$ under a schedule $W_C$ as
\begin{align}
 \mathbf{m}^C_W(t) = M^C[e^{C}_{W}(t) \equiv W_C].
\end{align}
A metric $d^C(\mathbf{x}, \mathbf{y})$ is then defined to study the model fidelity on a given context $C$.

Having established the above terminologies, for a model $M^{C}_{\epsilon}$ in a given context $C$ to be an $\epsilon$-mirror, it must satisfy the equivalence:
\begin{align}
    d^C(\mathbf{m}^C(t), \mathbf{r}^C(t)) \le \epsilon  \quad \textrm{for all } {D_\text{test}}.
\end{align}
This model $M^{C}_{\epsilon}$ is fit for purpose if and only if it is an $\epsilon$-mirror for $C$ and $\epsilon \le \epsilon_r$, where $\epsilon_r$ is a critical threshold based on context or engineering requirements.

Worden et al.~\cite{Worden2020} further augmented their proposed physics-based framework with a data-driven approach---a hybrid approach such as a machine learning model trained on $D_\text{train}$ to give a new model $M^{hC} (D_\text{train})$---as a means to meeting model updates requirements. In this case, the probabilistic nature of most machine learning models thus necessitated the need to assess model accuracy and as such to quantify uncertainty.

\subsection{Field Specific Adoption: SHM + DT}
\label{subsec: 6.6}
Following conceptual overviews on SHM and DT development, we now report applications of DT for SHM across two categories. As presented below, the aforementioned techniques are integrated as enabling technologies for the development of DT for SHM. 

\subsubsection{Predictive Diagnostic SHM}
We notice that most existing developments of DT for SHM purposes were aimed at realizing the real-time condition assessment of structures, and predictions of damage presence, type, and locations. We classify such an application of DT as predictive diagnostic SHM; some of which are reviewed below. 

Febrianto et al \cite{Febrianto2022} developed a digital twin using a statistical finite element model (StatFEM) for autonomous continuous monitoring and assessment of the structure. The authors employed hybrid simulation techniques for prediction. Fiber Bragg grating sensors were utilized for the continuous monitoring of a railway bridge. The measured data was the strain distribution of the bridge, which was previously collected from the point of construction to two years post-construction and validated with a deterministic FE model by other researchers. Their proposed framework accounted for uncertainties in sensor readings, applied loading, FE model misspecification errors, and train weight by making the FE component a random variable represented by a Gaussian process. The prior density of the FE component was obtained by solving a traditional probabilistic forward problem, and the posterior densities were inferred via the Bayes rule. The results revealed reasonable accuracy between measured and predicted data post-calibration of the DT, even in regions with missing sensor data.

Angjeliu et al \cite{Angjeliu2020} proposed a DT for the structural integrity assessment of a historic masonry Cathedral in Milan. The authors adopted a non-linear finite element model, based on data from photogrammetric geometric measurements, extensive in situ survey, and archive data. In addition, due to the historical nature of the building, there was no available structural model, thus the researchers employed a mixed modeling technique consisting of parametric modeling and computer-aided engineering (CAE) for the as-built model, which was then discretized to produce the FE model for DT development. Accelerometers were employed as sensors and placed on the nave of the cathedral (the target location), for data acquisition and model updates. Their proposed DT helped in understanding the force distribution and interdependencies between structural members, which was successfully utilized for damage detection and prediction within certain parts of the structure. In a similar trend of DT development for legacy buildings, Pregnolato et al \cite{Pregnolato2022} proposed a DT that was utilized for the predictive maintenance of the Clifton suspension bridge in Bristol, United Kingdom. In this case, three types of sensors including temperature sensors, displacement transducers, and strain gauges were utilized for data acquisition for model development and acquisition, and with the global aim of estimating the coefficient of friction of the tower saddles. The 3D FE model was developed in MIDAS Gen, which was calibrated with the previously developed models. The data interface was in two layers: a software layer, where model updating occurs, and the application layer which interprets the inputs and provides outputs. We highlight that these particular applications of DT for SHM utilize existing technologies previously introduced in other sections such as parametric BIM and sensing, thus suggesting their synchronizations for context-specific applications.

In tandem with our adopted concept on SHM-based DT, Ritto et al.~\cite{Ritto2021} developed a DT for dynamical systems. The authors emphasized the computational model, uncertainty quantification, and model calibration as the core of their DT. A prismatic bar with similar structural behavior was adopted as a theoretical physical asset. Their computational model was constructed using a 6-degree-of-freedom lumped parameter description and the uncertainty quantification component was accounted for by modeling parameters as uniform random variables. The authors trained and tested various data classifiers for appropriate model selection. Finally, they performed a series of analyses on their DT by varying parameters in their data sets, expressing several causal and resultant relationships between their DT components and data parameters. Their DT model was tested for damage detection and prediction but excluded model discrepancy and prognostic SHM.

\subsubsection{Predictive Diagnostic + Prognostic SHM}
For complete SHM processes, some researchers have developed DT for prognostic evaluations. By prognosis evaluation, we imply a time evolution study of a physical twin through its DT. This goes beyond damage prediction based on anticipated events but also includes understanding damage evolution such as crack propagation. With such possibilities, DT features such as planning, automated preventive maintenance, and efficient model updates can be realized \cite{Bordegoni2023}.

The models reviewed in \Cref{subsec: 6.5} are applied to particularly enhance SHM capabilities for both diagnosis and prognosis.
Specifically, the PGM developed by \cite{Kapteyn2021} was applied by \cite{Torzoni2024} to develop a DT framework for civil engineering structures. Torzoni et al.~\cite{Torzoni2024} presented three key steps for predictive digital twins using physics-based models and machine learning. First is the use of the PGM framework for predictive DT. For this, the observed data are assimilated using DL models as an initial estimate of the digital state. The structural health parameters that comprise the digital state are updated using sequential Bayesian inference based on health-dependent control policy to capture the structural health evolution. The second step is a simulation-based strategy to generate training datasets for the DL models i.e., numerical simulations of physics-based models were utilized to generate the datasets. For civil structures, the detection of damage patterns characterized by a small evolution rate remains the focus, prompting the use of stiffness as the ideal parameter. Reduced order modeling (ROM) based on proper orthogonal decomposition (POD) and Galerkin projection were employed to simplify the problem and save computational costs during the synthetic data generation. The third step comprises the assimilation of the populated datasets using two DL models. For every new observational data acquired, a classification task is performed to identify damage detection/localization. For every identified damage, the observational data are further processed with regression models to quantify the amount of damage at each damageable region. The developed methodologies were applied for two numerical experiments which comprised a simulated L-shaped cantilever beam and railway bridge. The results showed promising prospects for the application of DT for SHM under varying operational conditions, with low uncertainties. Also the framework as presented by \cite{Kapteyn2021} propels planning by suggesting suitable control inputs for future occurrences.   

Although not in civil engineering, Karve et al.~\cite{Karve2020} extended the diagnosis and prognosis aspect of SHM to include intelligent mission planning for unmanned aerial vehicles (UAV) to ascertain safe and reliable mechanical systems with maintenance-free operation. For mission planning optimization, the authors utilized load profile optimization to minimize the damage growth without compromising the level of work done. They specifically employed a hybrid strategy where the initial missions aim to minimize the final damage value, and later missions minimize the probability of exceeding critical crack size. With all these in place, damage-tolerant mission planning under uncertainty using DT was achieved.

\subsection{Remarks on DT and SHM}
We introduced the necessary arsenals to develop DT for SHM purposes. These include the fundamental processes of data acquisition through sensing technologies, data processing such as feature extraction via ML methods, and data interpretation from extracted features or raw data using computer vision and DL methods. Furthermore, the conventional SHM techniques that have been reportedly successful for diagnostic purposes were extended to include DT as a basis for enabling feedback and control, and ultimately achieving real-time SHM that spans both diagnostic and prognostic evaluations.

As a starting point for our remarks, we remind readers that SHM is one of the possible applications of the DT of a corresponding structural system. Thus, DT should not be viewed as a tool for SHM but rather as a system with SHM capabilities. For most existing DT-based SHM, the current capability level stops at real-time damage detection and prediction for better decision-making in terms of early warnings or precise maintenance. However, an ideal DT-based SHM should not only include damage detection, localization, and level of severity, but also provide information on damage propagation, life cycle estimation, preventive maintenance, and mission planning. From our review, we notice limited developments of DT to achieve such possibilities which we categorized as predictive diagnostic + prognostic SHM. Taking a closer look at this category, we notice that mission planning can be linked to the aviation industry such as the work done by Karve et al.~\cite{Karve2020} and Kapteyn et al.~\cite{Kapteyn2021}. This can be correspondingly achieved for civil structures in terms of automated predictive maintenance that will identify remedial actions for anticipated damage based on time evolution studies. In fact, this is requisite for achieving a level-5 DT.

For DT development, two techniques have proved important. First is the inevitable need for uncertainty quantification. Kapteyn et al.~\cite{Kapteyn2021} in their mathematical abstractions and corresponding PGM framework emphasized uncertainty quantification for all six parameters. Ritto et al.~\cite{Ritto2021} added a layer of uncertainty quantification in the computational model of a DT, and Karve et al.~\cite{Karve2020}  mandated uncertainty quantification for diagnostic, prognostic SHM, and also load-profile optimization for mission planning. We agree with the body of literature in this regard and particularly state that predictive SHM at every level can be viewed as a probabilistic SHM. As already established, the model update goes hand in hand with uncertainty quantification but not necessarily uncertainty quantification. As the name implies, model updating is a must for an accurate depiction of the digital states of the corresponding physical states of an asset. While sequential Bayesian inference is commonly employed, the need for continuous model updates can be computationally expensive due to the repetitive need for evaluations \cite{Zhu2020}. Also, the unique nature of civil engineering systems with very slow structural alteration and false anomalies from ambient excitations questions the need for all-time model updates. As proposed by Zhu et al.~\cite{Zhu2020}, pre-model update checks could be developed to ascertain the appropriate time for model updates. Pan et al.~\cite{Pan2021}  also demonstrated such an approach for a BIM-data mining integrated DT. We thus suggest more research on the most efficient approach for model updates for the DT development of structural systems. For both uncertainty quantification and model updates, randomization of parameters for probabilistic evaluations is critical. 

\section{Concluding Remarks}
\label{sec: 7}

The unconsolidated adoption of DT across several fields threatens its perception as a technological buzzword, which could easily perpetuate the civil engineering field if one looks at DT's level of development for its projects. Through the literature surveyed, we showed DT to be an advancement to existing technologies that make use of the great possibilities that currently exist in computational science and engineering. Notably, we showed DT to be an enabler for mission planning optimization in the aviation industry and as a platform for realizing the factory of the future in the manufacturing industry. Although we conceded to the slow adoption of DT in the civil engineering field, we equally presented several motivating examples of at least level-3 DTs developments for SHM. 

Specific to this work, we view DT as a viable solution to the unique challenges at the separate phases of civil engineering projects. This is in tandem with the fundamental concept of DT as a continuous technology across the entire life cycle of projects or products. We started by emphasizing that a complete DT which we termed level-5 DT should be capable of digital representation, real-time monitoring, prediction, interaction, and autonomous control. Beyond a service-based notion of DT, we noted the fundamental components of DT from a developmental angle. In this regard, we strongly stated that the distinguishing feature of a DT comprises: a physical asset, digital asset, real-time bidirectional data exchange, feedback, and control. However, most attempts to develop DT exclude the controlling or actuating capabilities of DT. 

The majority of our review focused on the phase-based development of DT to advocate for a holistic adoption of the same in the AEC industry. This should not be confused with a proposition for a singular DT for the entire civil engineering phases. For this reason, we presented DT as a context-based representation of a real asset to avail accurate and computationally tractable models. Furthermore, we adopted self-sufficient narrations of the conventional techniques at each phase as a basis for DT development and applications. We equally provided remarks specific to each phase to provide insights on the forward-thinking approach for DT development and adoption for civil engineering projects. For example at the planning/design phase, we highlighted the realization of indirect DT using similar or historic structures for a more informed design of high-risk projects. In the construction phase, we showed the role of BIM, sensing, and IoT in developing highly sensitive DT that are dynamic enough to capture the transient nature of projects at this phase. We then showed the promise of DT for robust SHM built on several technological and mathematical concepts---such as model update and computer vision--- to make prognostic SHM commonplace in civil engineering. Another important insight includes the synchronization of the dominant technologies across the different phases, one which was apparent in the reviews presented on BIM, sensing, and computer vision.

Although not detailed in this work, other important technologies such as cloud computing and cyber-physical systems are important in the framework of DT development. Aside from a technological view of DT, there exists an important cultural view. For this, the place of data privacy is highly contentious and the role of DT for the civil engineering workforce must be equally explored. If consciously adopted, DT will become a competitive tool for sustainable and resilient civil engineering practice amidst the mounting environmental and social challenges in the AEC profession. 

\section*{Acknowledgment} %

The authors thank the University of Houston Division of Research for providing startup fund (Award No.~000186084) to support this research.

\bibliographystyle{asmejour}   %

\bibliography{DT-civil} %

\clearpage

\end{document}